\documentclass[journal=nalefd,manuscript=letter]{achemso}

\usepackage[version=3]{mhchem} 
\usepackage{siunitx}             
\usepackage{hyperref}

\newcommand{\MoS}[1]{\ce{MoS#1}}

\author{Yury Polyachenko}
\email{polyachenkoya@princeton.edu}
\affiliation[PPPL]{Princeton Plasma Physics Laboratory, Princeton, NJ 08540, USA}
\alsoaffiliation[Princeton]{Department of Chemistry, Princeton University, Princeton, NJ 08544, USA}

\author{Yuri Barsukov}
\alsoaffiliation[PPPL]{Princeton Plasma Physics Laboratory, Princeton, NJ 08540, USA}

\author{Shoaib Khalid}
\alsoaffiliation[PPPL]{Princeton Plasma Physics Laboratory, Princeton, NJ 08540, USA}

\author{Igor Kaganovich}
\email{ikaganov@pppl.gov}
\alsoaffiliation[PPPL]{Princeton Plasma Physics Laboratory, Princeton, NJ 08540, USA}

\title{Transition Metal Dichalcogenide \MoS2: oxygen and fluorine functionalization for selective plasma processing}

\newcommand{\Esputt}{E_{\text{sputt}}}
\newcommand{\EsputtS}{E_{\text{sputt,S}}}
\newcommand{\EsputtMin}{E_{\text{sputt,min}}}
\newcommand{\Eheadon}{E_\text{head-on}}
\newcommand{\Ecollision}{E_\text{collision}}
\newcommand{\Eescape}{E_\text{escape}}
\newcommand{\Ehit}{E_\text{hit}}
\newcommand{\Nhits}{N_{\text{hits}}}
\newcommand{\Tmax}{T_\text{max}}
\newcommand{\Tmelt}{T_\text{melt}}
\newcommand{\Ti}{T_\text{i}}
\newcommand{\vAr}{v_\text{Ar}}
\newcommand{\vCoM}{v_\text{CoM}}
\newcommand{\mS}{m_\text{S}}
\newcommand{\mO}{m_\text{O}}
\newcommand{\mX}{m_\text{X}}
\newcommand{\mAr}{m_\text{Ar}}
\newcommand{\mI}{m_\text{I}}
\newcommand{\EAr}{E_\text{Ar}}
\newcommand{\EArPerp}{E_{\text{Ar} \perp}}
\newcommand{\EArX}{E_\text{Ar-X}}

\newcommand{\Psputt}{P_{\text{sputt}}}
\newcommand{\thtInherent}{\theta_\text{inherent}}
\newcommand{\thtThr}{\theta_{\text{thr}}}
\newcommand{\thtOpt}{\theta_{\text{opt}}}
\newcommand{\thtArX}{\theta_\text{Ar-X}}
\newcommand{\phiOpt}{\varphi_{\text{opt}}}
\newcommand{\rArX}{r_\text{Ar-X}}
\newcommand{\rArO}{r_\text{Ar-O}}
\newcommand{\rWell}{r_\text{well}}
\newcommand{\rHardsphere}{r_\text{hard-sphere}}
\newcommand{\UArO}{U_\text{Ar-O}}
\newcommand{\UArX}{U_\text{Ar-X}}
\newcommand{\UinvArX}{U^{-1}_\text{Ar-X}}
\newcommand{\dAtomX}{d_\text{atom,X}}

\begin{document}


\begin{figure}
  \centering
  \includegraphics[width=0.6\linewidth]{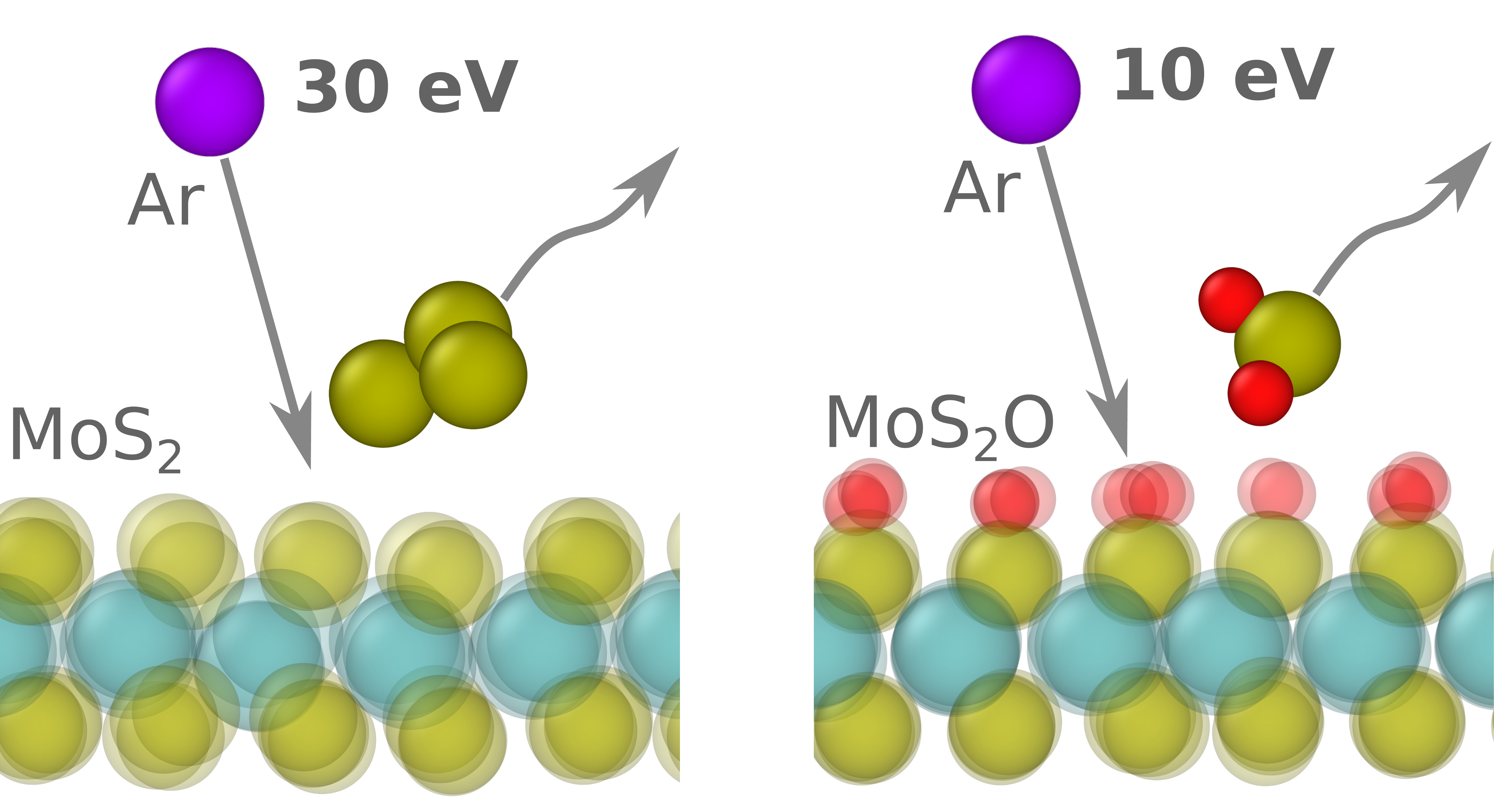}
\end{figure}

\clearpage

\begin{abstract}

Low-temperature plasma processing is a promising technique for tailoring transition metal dichalcogenides (TMDs). For chalcogen substitution processing, a key challenge is to identify the ion energy window that enables selective chalcogen removal while preserving the metal lattice. Using ab-initio molecular dynamics (AIMD), we demonstrate that oxygen and fluorine functionalization widen the processing window by significantly lowering the sulfur sputtering energy threshold ($\EsputtS$) of \MoS2 from $\sim 30$ eV to $\sim 10$ eV via formation of sputtering products such as \ce{SO2} and \ce{SF_n}. Additionally, we show that experimentally relevant cryogenic temperatures strongly affect $\EsputtS(T)$. The dependence is confirmed via AIMD and also predicted by a mechanistic parameter-free theory, suggesting that $\Esputt(T)$ generalizes to other TMDs, functionalization, and surface impacts in general. Our results highlight oxygen/fluorine functionalization, ionic impact angle, and material temperature to be key control parameters for selective, damage-controlled chalcogen removal in TMD processing.

\end{abstract}



Plasma processing offers a powerful technique for tailoring material structure and properties\cite{Nastasi1996}. Plasma processing has been widely applied to two-dimensional (2D) materials, including graphene\cite{Nanda2017, Lemme2009, Archanjo2012} and transition metal dichalcogenides (TMDs)\cite{Wang2016, Stanford2017, Bertolazzi2017, Chow2015, Lin2016, Fox2015, Ma2017}. For TMD processing, ion bombardment underpins a variety of processing steps such as etching, cleaning, and doping\cite{Sovizi2023} where selective removal of targeted atoms is required while preserving the lattice structure. Among TMDs, monolayer \MoS2 in its 2H phase stands out as a particularly promising candidate owing to its direct band gap, comparable to silicon, but at a much smaller thickness\cite{khalid2024deep}. A critical factor in plasma processing of \MoS2 is identifying the ion energy window that enables selective chalcogen removal while preserving the metal lattice.

Pristine \MoS2 has been widely studied through both experimental and theoretical approaches: precise ionic bombardment \cite{Lu2020, Ma2013, Gutirrez2012, Wang2023, Bae2017, Parkin2016, Park2024}, excitations \cite{Komsa2012, Kretschmer2020, Yoshimura2023, Novko2016, Palummo2015, Bai2025, Shi2013}, multi-charged ions \cite{Kozubek2019, Grossek2022, Niggas2022, Skopinski2023}, ion neutralization \cite{Masuda1993, Buitrago2024, Gainullin2020}, substrate effects \cite{Ghaderzadeh2020, Kretschmer2018, Wu2021}, strain \cite{Bertolazzi2011, Albaridy2023, Frisenda2017, Wang2018}. Experiments have shown that \ce{Ar+} ions with kinetic energies of approximately 50 eV are sufficient to generate sulfur vacancies (S vacancies), while energies near 100 eV are required to remove Mo atoms and ultimately etch the entire layer in \MoS2\cite{Lu2020}. This establishes a baseline energy window for plasma processing. However, modern techniques are capable of creating ions with temperatures as low as $\Ti \sim 1$ eV and also applying a bias of $\sim $ 10-20 eV \cite{Chopra2024, Son2025}. This allows controlling ion energies of $10-20$ eV to within $\pm$1 eV. Therefore, these plasmas can be used to create an ion flux with precisely controlled energies. In this work, we calculate the sputtering threshold energy, $\EsputtS$, of \MoS2 with significantly improved accuracy and propose a novel strategy to reduce this threshold by functionalizing the TMD surface with oxygen or fluorine. Based on ab-initio molecular dynamics (AIMD) simulations, we develop a sputtering mechanism that explains the observed results, including a prediction of temperature dependence of $\EsputtS$ for oxygen-functionalized \MoS2. 


Previous theoretical studies have estimated that to eject a S atom from \ce{MoS2}, approximately 7 eV of energy directed exactly outward from the material needs to be transferred to the sulfur atom \cite{Kretschmer2020-io, Komsa2012}. This value decreases to about 4–6 eV when considering excitations of various valence electrons \cite{Bai2025}, and can be as low as 2.2 eV in the presence of multiple excitations of core electrons, i.e. core-electron excitations. Electron scattering experiments have inferred an effective sulfur desorption energy of approximately 1.5 eV \cite{Kretschmer2020}, suggesting that incident electrons generate multiple deep excitations while transferring momentum to the S atom. Low-energy plasma processing is generally conducted with ion energies not exceeding approximately 100 eV\cite{Sovizi2023}. Under these conditions, the generation of deep excitations is unlikely, since the sulfur 2$p$ peaks in XPS spectra of \ce{MoS2}, corresponding to core excitation energies, appear above 150 eV \cite{Jones2022}. Moreover, coupling between electronic and ionic motion is inefficient due to their large mass disparity, which makes fast energy transfer to electrons infeasible and gives the ion time to lose its energy in multiple collisions. Therefore, we estimate that the lower bound of the sulfur escape energy threshold is $\Eescape = 6–7$ eV. 

During TMD-ion collisions, the energy $\Eescape$ is what needs to be transferred to the impacted S atom. This transferred energy is smaller than the energy of the incoming projectile $\Ehit$ due to the mass ratio $\mS / \mI \neq 1$ between the impacted S atom and the ion. Therefore, the lower bound for the sputtering projectile energy $\Ehit$ is larger than the escape energy by the following factor:

\begin{equation} \label{eq:EkinTransf}
\Ehit = \Eescape \cdot \left( \frac{\sqrt{\mS/\mI} + \sqrt{\mI/\mS}}{2} \right)^2.
\end{equation}

However, the collisions are not binary and involve other atoms. Thus, the energy transferred from ion to sulfur atoms can be shared with neighboring molybdenum or other sulfur atoms, so the threshold energy for removal can be higher.

Earlier molecular dynamics simulations on 2D materials have demonstrated that when ion energies exceed 100 eV, the sputtering yield is mainly controlled by secondary collisions involving atoms that are reflected or knocked out from the supporting substrate \cite{Kretschmer2018}. However, at lower energies close to the damage threshold (around 10 eV), the sputtering process is expected to be substantially different. The energy barely sufficient to cause a sulfur ejection is not enough to significantly perturb the material below the top TMD layer, especially given relatively wide gaps between TMD layers. Thus, the substrate can be safely excluded from the simulations. Instead, so‑called "chemically enhanced physical sputtering" was described for silicon, when ejecting more chemically stable reaction products rather than individual atoms often reduced the minimum energy required for sputtering \cite{RamanaMurty1995}. When an ion collides with the top sulfur atom, the impacted sulfur atom initiates a cascade of intermediate collisions that can affect TMD lattice. Careful consideration needs to be given to these intermediate collisions so that the TMD lattice is restored after the collision and is not damaged. 

In this research, we propose a two-step chemically-enhanced physical sputtering mechanism that minimizes the sputtering threshold energy $\EsputtS$ of \ce{MoS2} functionalized by oxygen or fluorine:

\begin{enumerate}
    \item The \ce{Ar} impact induces atomic rearrangements that facilitate the formation of (meta)stable gas-phase species such as \ce{SO2}, \ce{SF4}.
    \item Under the above condition, efficient momentum transfer from the incoming \ce{Ar} to the ejected molecular species is achieved by optimizing the number and directions of intermediate collisions that lead to desorption of these products.
\end{enumerate}

\begin{figure}
  \centering
  \includegraphics[width=1\linewidth]{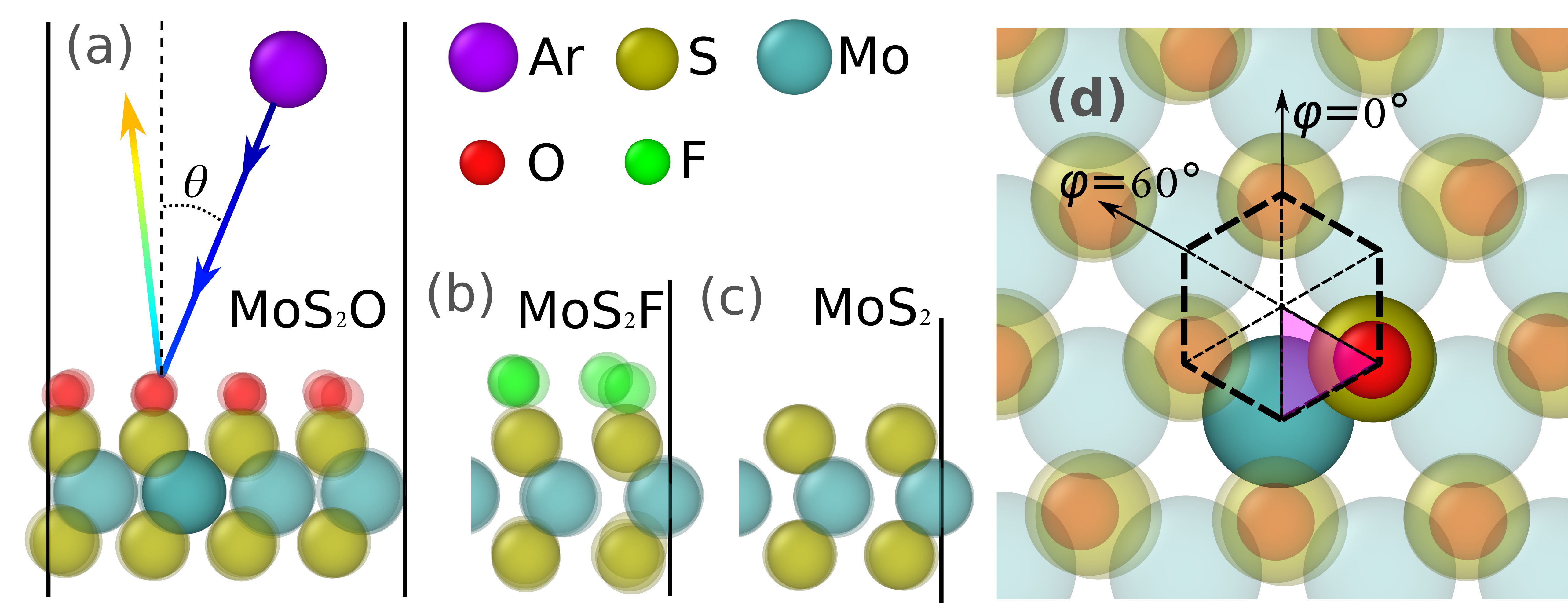}
  \caption{Simulation setup schematic. (a): Side view of the full \ce{MoS2O} equilibrated supercell (4x4) initial state. \ce{Ar} atom's trajectory is shown by a rainbow line with color representing time from blue to red. (b,c): Representative parts of the \ce{MoS2F} and \ce{MoS2} initial equilibrated states (side view), respectively. The irregularity of the fluorine atoms' positions is discussed in the main text. (d): Top view of \ce{MoS2O}. The purple shade triangle shows the minimal set of hit-points that represent all other points by symmetry. The in-plane far-field approach angle, $\varphi$, shows its origin $\varphi = 0^{\circ}$ and its positive direction $\varphi = 60^{\circ}$. The hit angle remains very close to the initial far-approach angle in cases considered in this study.
  }
  \label{fig:F1}
\end{figure}

The simulation setup for collisions of the incoming \ce{Ar} with monolayer \ce{MoS2} is shown in Figure \ref{fig:F1}.

\begin{figure}
  \centering
  \includegraphics[width=0.7\linewidth]{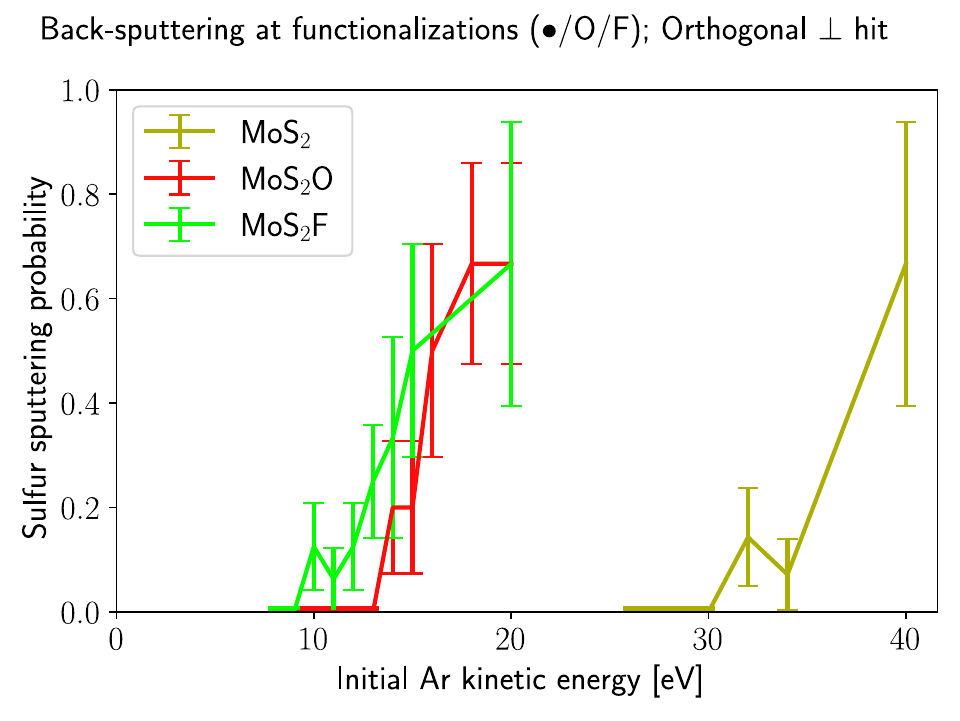}
  \caption{ Probability to eject sulfur from pristine and functionalized \ce{MoS2} back into the half-space the impact came from. Functionalization with O and F substantially lowers the sputtering threshold energy, indicating enhanced chalcogen removal efficiency at reduced \ce{Ar} energies. For each of three TMDs, the impact point was optimized to minimize $\EsputtS$ (details in Figure \ref{fig:F2S5}). Materials were equilibrated at 116 K.
  }
  \label{fig:F2}
\end{figure}

In this paper, we show that oxygen or fluorine functionalization reduces the sputtering threshold energy $\EsputtS$ in \ce{MoS2}, as illustrated in Figure \ref{fig:F2}. This phenomenon is explained by the observation that small (meta)stable molecular fragments, such as \ce{SO2} and \ce{SF4}, can form more easily from a few TMD atoms actively involved in a low-energy \ce{Ar} impact, compared to the ejection of pure sulfur atoms or clusters (\ce{S2}, \ce{S6}, or \ce{S8}). The enhanced formation of these O- or F-containing fragments can be attributed to O/F's high electronegativity, which stabilizes intermediate fragments relative to the rest of the material and facilitates their ejection under lower-energy impacts. These findings suggest that selective functionalization provides a practical route to control the sputtering behavior of different TMDs at energies near the damage threshold.

As a preliminary step, we examined the feasibility of O and F adsorption on \ce{MoS2} before analyzing the impacts of O and F covered surfaces. It is well known that oxygen exhibits strong reactivity with TMDs \cite{Xie2013, Nan2014, Kang2014, Chow2015ext, Yang2013, Islam2014, Yang2015, KC2015, Zhou2013, Pet2018, Giannazzo2017, Reidy2023}. For example, even molecular \ce{O2} has been shown to fill S vacancies in \ce{MoS2} with an activation barrier of approximately 1.05 eV \cite{Nan2014}. Previous calculations have further indicated that \ce{O2} can dissociate into atomic oxygen, and, subsequently, adsorb onto \ce{MoS2}, with a total barrier of about 1.6 eV \cite{KC2015}. Isolated atomic oxygen was also reported to adsorb on \ce{MoS2} with a binding energy of 0.87–1.12 eV, depending on the surface coverage \cite{KC2015}. Our calculated adsorption energy for O on \ce{MoS2} is in close agreement with the known results, as shown in Figure \ref{fig:F1S1}b of the supplementary information (SI). Differences of $\sim 0.1-0.2$ eV may have resulted from a smaller unit-cell and k-point sampling, but are deemed acceptable given the studied energy scale of $\sim 10$ eV. Furthermore, we performed meta-dynamics calculations to verify that our conclusions based on the potential energy surfaces (PES) qualitatively hold for the corresponding free-energy surfaces (FES) as shown in Figure \ref{fig:F1S1}d and \ref{fig:F1S1}e.

Oxygen has been experimentally shown to play a significant role in the thermal etching of TMDs, initiating around $345^{\circ}\mathrm{C}$ \cite{Zhou2013}. Under ambient conditions, approximately month-long exposure to molecular \ce{O2} also induces sulfur vacancies \cite{Pet2018}. In both cases, the formation of stable gas-phase species such as \ce{SO2} has been proposed as a key step in oxygen-assisted etching. On the other hand, it has been demonstrated that TMDs can undergo oxygen functionalization, which is detectable in characterization techniques such as PL and XPS \cite{Sovizi2023}. Finally, processing energy windows were investigated\cite{Lu2020}.

Despite the damage energy threshold being a key parameter in processing, the interplay between TMD functionalization and changes in the sputtering energy threshold has not yet been quantitatively explored. Inspired by observations above, we show a significant drop in the threshold sputtering energy induced by adsorbed oxygen and fluorine, which can considerably ease chalcogen plasma-assisted removal in TMDs. We quantify how such functionalization could extend the \ce{Ar} ion energy window for plasma processing and enhance the selective removal of chalcogen atoms while preserving the underlying metal lattice. We also confirmed and explained coupling of the sputtering threshold to the impact angle and to TMD temperature. Reported effects rely on a simple mechanistic theory and are expected to generalize to other TMDs and appropriate functionalizations.


We first investigate impacts orthogonal to the material plane ($\theta = 0$, see Figure \ref{fig:F1}a). Figure \ref{fig:F2} shows the probabilities of sulfur ejection $\Psputt$ for pristine and functionalized \ce{MoS2}. Atomic oxygen functionalization decreases $\EsputtS$ from $\sim (31 \pm 1)$ eV to $\sim (14.0 \pm 1)$ eV. Atomic fluorine is even more effective with $\EsputtS\sim (9.5 \pm 0.5)$ eV. \textcolor{black}{Full surface coverage is considered in our model. The reduction in the sputtering threshold arises from the formation and desorption of products such as \ce{SO2} and \ce{SF4}. Thus, using combinatorial arguments about occupation of neighboring sites by oxygen or fluorine, we show that surface coverage $c \in (0;1)$ should decrease the probability of the proposed mechanisms by $c$ (details in SI ``Surface coverage fraction effects'').} No back-sputtering of \ce{Mo} atoms is observed up to $\EAr = 50$ eV for \ce{MoS2} and the layer destruction is seen at 100 eV, confirming the experimentally known constraint \cite{Lu2020}. No Mo back-sputtering is seen below 50 eV for \ce{MoS2O} and \ce{MoS2F}, so the processing energy window indeed seems to be widened by O/F functionalization. 

These values correspond to the energy of an Ar projectile that results in sulfur sputtering at least once out of $\Nhits$ given the most susceptible hit-point (detailed in Figure \ref{fig:F2S5}). At least $\Nhits \geq 14$ was used in all reported simulations. This describes sampling at a fixed hit-point, and we estimate it to correspond to sampling $\sim 4000$ impacts per elementary triangle (see SI). The threshold definition technically depends on $\Nhits$, which is expected since any event can be observed once given sufficient sampling. However, we estimate (details in SI) that the dependence is weak $\sim \sqrt{\ln(\Nhits)}$. We also estimate a lower bound for sputtering yield near the computed energy threshold $Y_S \geq 0.0036 \cdot \Psputt$ (detailed in SI), which is approximately consistent with force-field MD results for \ce{MoS2} \cite{Kretschmer2018}. Additionally, sputtering requires defining the timescale of ejection because weakly adsorbed products such as \ce{SO2} can take a long time to desorb. These effects set an energy range $\sim$ 1 eV wide near the energy threshold, where increasing observation time meaningfully decreases the calculated energy threshold. Based on representative longer AIMD (Figure \ref{fig:F2S4}), we choose 2 ps as the cutoff time that optimally separates unimpeded ejections from thermal product desorption. Finally, the determination of the damage energy threshold instead of the direct yield requires only the simulation of impacts of damage-free material because we only quantify the transition from no-damage to some-damage. This allows us to avoid relatively long simulations of many consecutive impacts of the same cell and instead run many independent single impacts in parallel.

The energy error bars represent the spacing of the energy grid used to determine the sputtering threshold because statistical errors were found to be much smaller. The precise threshold values can change depending on the specific approximations used in DFT, such as the selected functional or pseudo-potential. For this reason, we did not further refine the energy grid. We demonstrate in SI that the results are sufficiently insensitive to the choice of DFT functional, dispersion correction scheme, k-point sampling, energy cutoff, and other computational settings. Because our study focuses on oxygen, we explicitly examine the relevant spin states. Oxygen is shown to have a singlet ground-state when bound to TMDs (Figure \ref{fig:F1S1}b), and the SI outlines our reasons for running all collision dynamics in fixed singlet spin state. For completeness, the triplet is shown to qualitatively preserve the significant drop in the sputtering energy threshold $\EsputtS$ from O/F functionalization. We note that the qualitative relative effect of the substantial reduction in $\EsputtS$ is expected to be insensitive to most of the DFT details, as it arises from clear physical mechanisms discussed below.

These results suggest that O and F pre-functionalization can effectively expand the ionic energy window for plasma processing, allowing selective chalcogen removal at lower projectile energies, while minimizing damage to the underlying metal scaffold.

\begin{figure}
  \centering
  \includegraphics[width=1\linewidth]{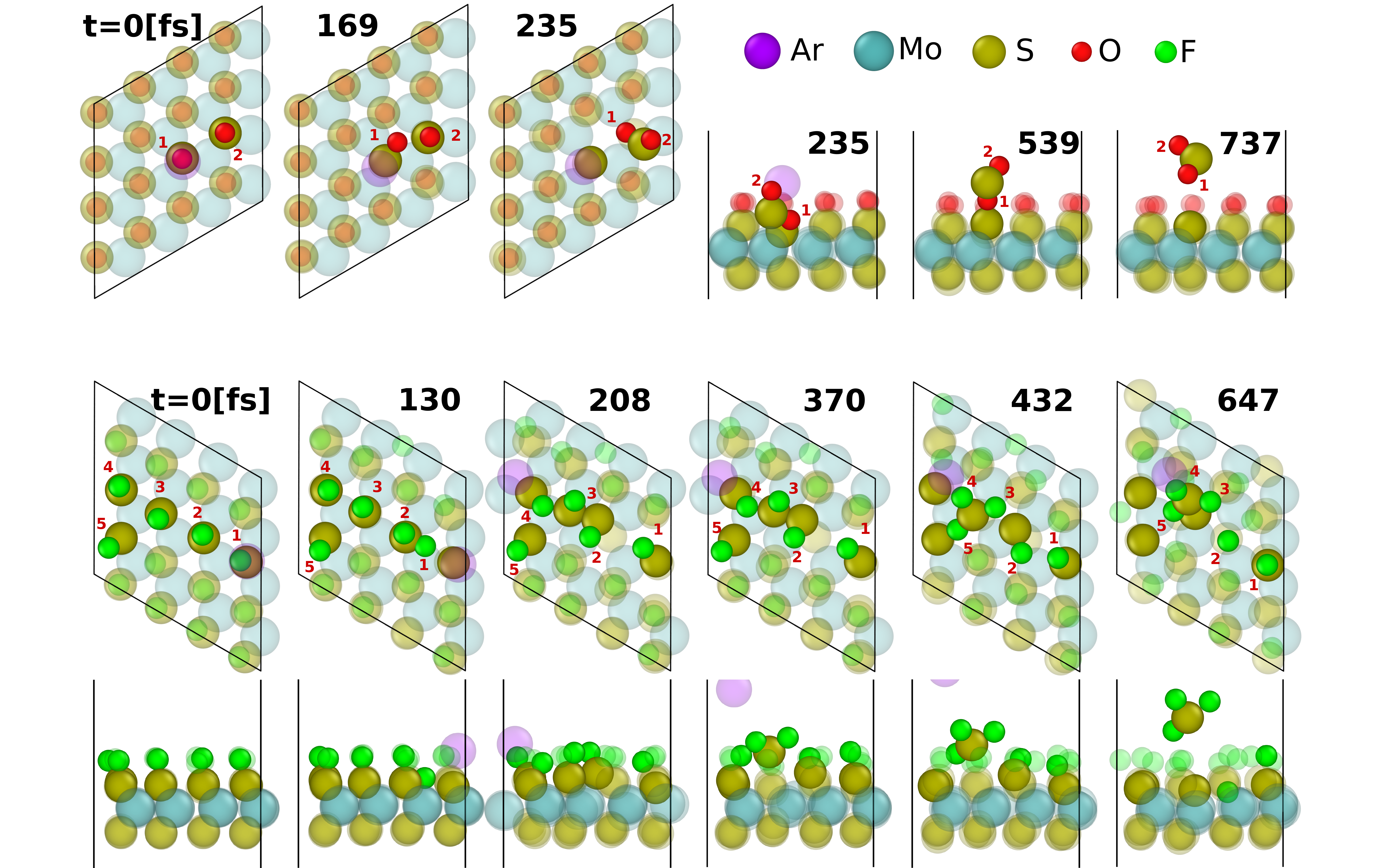}
  \caption{(Top): A typical orthogonal \ce{Ar} collision with \ce{MoS2O}. Opaque atoms are those that ultimately play a major role in the collision. Gray numbers are timestamps in [fs]. The sputtering mechanism of \ce{MoS2O} that realizes at the lowest required projectile energy involves an O atom being pushed in the direction of the nearest S atom that already has another O bound to it, allowing them to combine and form \ce{SO2}, which then escapes. Small red numbers enumerate O atoms to help follow them during collisions. A collision at an angle is shown in video V1. (Middle and Bottom): A typical orthogonal \ce{Ar} collision with \ce{MoS2F}. Side and top views are both provided to illustrate more complex atom movements during the collisions. Small red numbers enumerate F atoms to help follow them during collisions. Both materials were equilibrated at 116 K, both impacts were with 15 eV.}
  \label{fig:F3}
\end{figure}

A qualitative understanding of why the threshold drops so significantly can be obtained by examining ejection events that are representative of \ce{MoS2O} and \ce{MoS2F}, as shown in Figures \ref{fig:F3} and \ref{fig:F3S1}.

A typical collision that results in sulfur sputtering from an O- or F-functionalized \ce{MoS2} involves the formation of (meta)stable products (such as \ce{SO2} or \ce{SF4}) that later escape. This is different from pristine \ce{MoS2} sputtering (details are provided in Figure \ref{fig:F3S1}), which first mechanically breaks bonds of S atoms that later may assemble into \ce{S_n} cluster products. Accordingly, we ascribe the substantial variation in sputtering energy thresholds to the distinct sputtering mechanisms involved.

A noticeable difference in typical collision events between \ce{MoS2O} and \ce{MoS2F} is attributed to the differences in equilibrium positions of adsorbed O and F atoms on the TMD surface, shown in Figure \ref{fig:F3} (at $t=0$). More details are discussed in the SI. 

The \ce{MoS2O} structure keeps all symmetries of \ce{MoS2} and oxygen atoms are adsorbed directly above the sulfur atoms (see Figure \ref{fig:F3} ($t=0$) and Figure \ref{fig:F2S2}). In contrast, \ce{MoS2F} has a more disordered F layer (see Figure \ref{fig:F3} at $t=0$ and Figure \ref{fig:F2S3}). \textcolor{black}{We believe this is analogous to ``Peierls distortions'' \cite{Peierls2001, Kagoshima1981, Burdett1983, You2021} due to odd number of electrons brought to a unit cell by a fluorine atom (see SI for more discussion)}. This leads to more chaotic collision dynamics. This is indicated by 5 F atoms significantly participating in the \ce{MoS2F} impact, as opposed to only 2 O atoms for the \ce{MoS2O} impact. Sputtering probabilities as a function of the ejected species for conditions of Figure \ref{fig:F2} are shown in Figure \ref{fig:F2S1}. It suggests that sputtering products from \ce{MoS2F} are more varied than those from \ce{MoS2O}, which further indicates more disorder in the \ce{MoS2F} case. A similar diversity of \ce{SF_n} products was reported before for \ce{F2} plasma \cite{Farigliano2023}.

\begin{figure}[h!]
  \centering
  \includegraphics[width=0.7\linewidth]{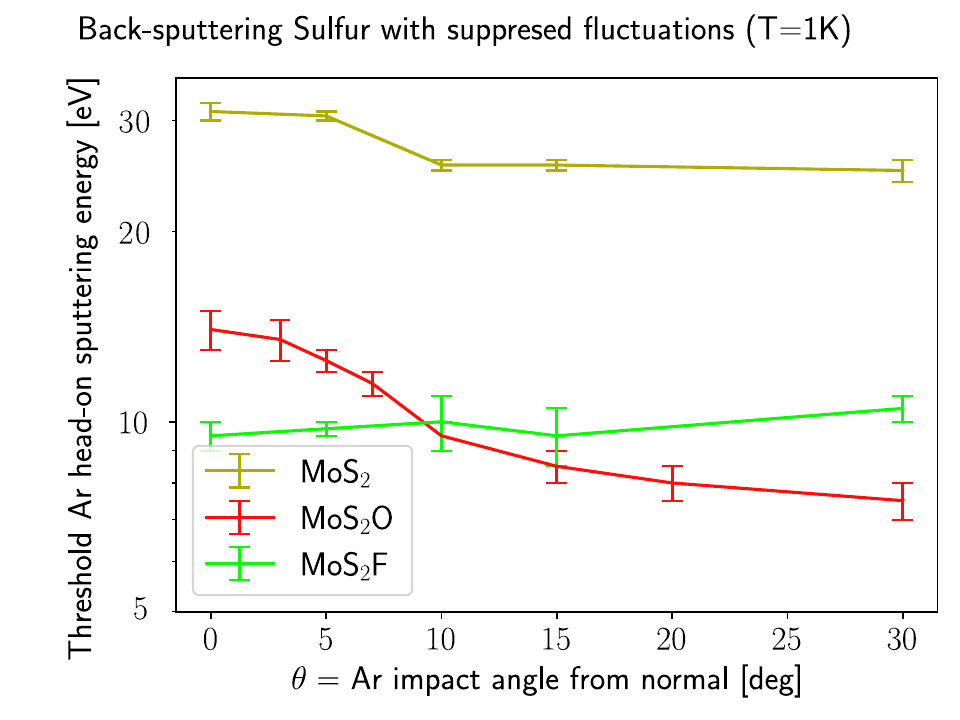}
  \caption{Angular dependence of the sputtering threshold energy $\EsputtS$ for \ce{MoS2} (dark yellow), \ce{MoS2F} (light green) and \ce{MoS2O} (red) for head-on impacts. Thermal fluctuations of target S/F/O atoms in the system are suppressed by formally setting $T = 1$K. Each curve is obtained after optimizing the impact point and the in-plane angle $\varphi$ to minimize $\EsputtS$, as shown in Figure \ref{fig:F4S2} for \ce{MoS2O}. Simulations at $\theta = 45^{\circ}$ were done for \ce{MoS2} and \ce{MoS2O} and they showed $E_{MoS_2}(45^{\circ}) > 35$ eV and $E_{MoS_2O}(45^{\circ}) > 14$ eV. Error bars reflect the step of the energy grid used to pin down the threshold.
  }
  \label{fig:F4}
\end{figure}

The observed difference in \ce{MoS2O} and \ce{MoS2F} ground states may be utilized for high-specificity etching due to the following: Because \ce{MoS2O} preserves the well-structured \ce{MoS2} lattice, collision results depend strongly on \ce{Ar} incident angle. Correspondingly, a strong angular dependence in the sputtering threshold was found for \ce{MoS2O} and \ce{MoS2} as shown in Figure \ref{fig:F4}. 

The threshold decreases significantly and non-monotonically with increasing impact angle, reaching about $(7.5 \pm 0.5) \text{eV} = \EsputtS(\theta = 30^{\circ}) \approx 0.54 \cdot \EsputtS(\theta = 0^{\circ})$ for the \ce{MoS2O}. The value $\EsputtS(\theta = 45^{\circ})$ (not shown on the plot) is $ > 14$ eV, so $\theta \approx 30^{\circ}$ is optimal for sputtering \ce{MoS2O}. A similar decrease by $\sim$ 6-7 eV is seen for pristine \ce{MoS2}, which also has a clear periodic structure before the impact.

These three presented TMD cases demonstrate a correlation between the degree of order in a system's ground state and its sensitivity to the angle of projectile impacts. Specifically, sufficiently disordered systems such as \ce{MoS2F} lose their sensitivity to impact direction.


To solidify our understanding of the proposed sputtering mechanisms, we formulate a simple theory that predicts the dependence of the sputtering threshold $\EsputtS$ on the TMD temperature $T$. The theory does not require MD simulations at different temperatures, but uses only the temperature-independent curve $\EsputtS(\theta)$ from Figure \ref{fig:F4} and simple collision models (see SI). We show below that such a simple and parameter-free theory is well correlated with MD simulation results, suggesting the proposed simple mechanisms indeed take place.

\begin{figure}[h!]
  \centering
  \includegraphics[width=1\linewidth]{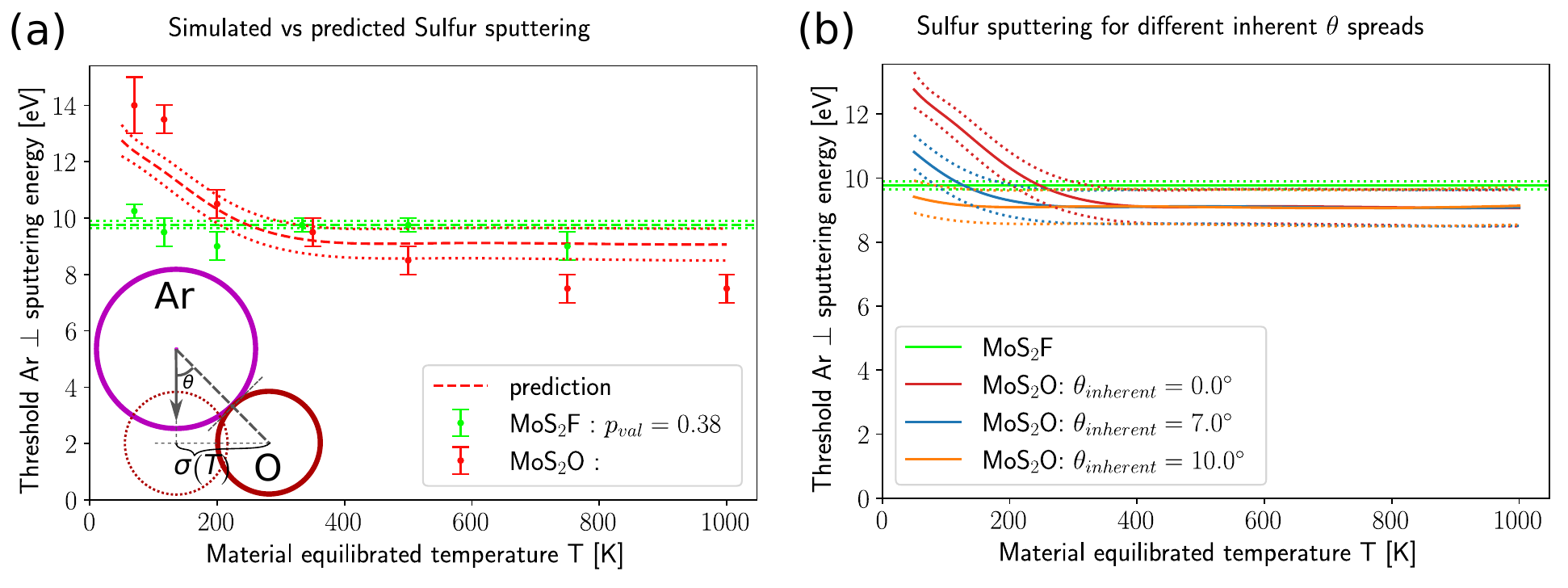}
  \caption{(a): AIMD simulation threshold (datapoints) at different temperatures for $\theta = 0$ and their predictions (dashed: mean, dotted: confidence interval) based on $\EArPerp(T)$ from eq.\eqref{eq:mainTh}. Red and green correspond to \ce{MoS2O} and \ce{MoS2F} respectively. No fitting parameters were used (with the exception of the spline parameters used for interpolating spline data of $\EArX(\theta)$ from Figure \ref{fig:F4}). (a-inset): A schematic of an \ce{Ar} impacting an O atom (red solid circle) that thermally fluctuated from its equilibrium position (red dashed circle). The \ce{Ar} velocity is directed exactly at the O equilibrium position, which was shown to be the most damage-susceptible point of \ce{MoS2O} in Figure \ref{fig:F2S5}. However, the impact is not head-on due to a thermal fluctuation of magnitude $\sigma(T)$ of the O atom. (b): Red, blue and yellow: Our predictions for $E_{\perp, sputt}(T)$ of \ce{MoS2O} for inherent $\theta$-spreads of the incoming \ce{Ar} of $0^{\circ}$, $7^{\circ}$, and $10^{\circ}$ respectively. Green: the $T$-independent $\EsputtS$ of \ce{MoS2F}. Dotted lines in (a) and (b) show confidence intervals based on error bars in Figure \ref{fig:F4}.
  }
  \label{fig:F5}
\end{figure}

The proposed theory is based on connecting the angular dependence $\EsputtS(\theta)$ and thermal fluctuations of target atoms, which cause an unavoidable spread in deflection angles. Because the magnitude of thermal fluctuations of target O/F atoms in the material (about 0.15 \AA{}, Figure \ref{fig:F4S1}) is comparable to the distances at which \ce{Ar} interacts strongly with the O/F atoms (i.e., hard-core repulsion interaction distance, about 1 \AA{}, see Figure \ref{fig:F4S1}), a certain range of deflection angles, $\theta_T$, often occurs due to thermal fluctuations. On the other hand, each \ce{Ar} impact energy $\EArPerp$ has a threshold angle $\thtThr(\EArPerp)$ such that impacts at larger angles $\theta > \thtThr(\EArPerp)$ result in sputtering. This means that if the thermal spread of deflection angles $\theta_T$ covers $\thtThr$ for a given temperature, i.e. $\theta_T(\EArPerp) > \thtThr(\EArPerp)$, then deflection angles sufficient for sputtering at the given $\EArPerp$ will be realized frequently, so sputtering will occur frequently. This is quantitatively explained in the SI in Figure \ref{fig:F4S0}. Quantification of this logic is given in the SI, and it yields the oxygen lines in Figure \ref{fig:F5}.

Notably, this is a parameter-free theory in the sense that no fitting to AIMD $\EsputtS(T)$ was done to obtain red dashed theory line in Figure \ref{fig:F5}a. Still, the predicted behavior is confirmed in AIMD simulations. It suggests the proposed sputtering mechanism and the theory that follows from it are reasonable and consistent approximations that provide insights into atomistic mechanisms of sputtering of \ce{MoS2} and its functionalized modifications. Therefore, the proposed sputtering picture can inform TMD simulations at the continuum level and potentially explain the nonlinearities associated with oxygen involvement \cite{Lee2016}.


Our analysis shows how functionalization of \ce{MoS2} with O or F can significantly lower the sputtering energy threshold from $\sim 30$ eV down to $\sim 10$ eV, as shown in Figure \ref{fig:F2}. Such functionalization may be used for TMD processing with minimal damage to the metal scaffold. Additionally, this type of functionalization enables spatial control over the etching process, which can be realized by selectively depositing oxygen or fluorine in accordance with a mask. Employing partial coverage with oxygen/fluorine can allow accurate control over the creation of sulfur vacancies by selectively sputtering a defined fraction of surface atoms that have been functionalized with oxygen/fluorine, and this approach is relatively insensitive to the exact plasma exposure time.

\textcolor{black}{The main mechanism responsible for the reduction in the energy threshold is attributed to the formation and subsequent desorption of products such as \ce{SO2} and \ce{SF_n}. A fully quantitative investigation is required to establish the limits of generalization of this mechanism; however, we expect similar behavior to occur in other TMDs and for other functionalizations. Preliminary AIMD simulations for \ce{MoSe2}, \ce{WS2}, and \ce{WSe2} with 100\% O and F termination indicate that the formation of species such as \ce{SF3}, \ce{SO2}, \ce{SeF3}, and \ce{SeO2} dominates the sputtering pathways near the threshold energies. Functionalization with O or F similarly reduces the sputtering threshold by approximately a factor of three, from $\sim$30 eV to $\sim$10 eV. In the case of oxygen termination, the hexagonal structure is largely preserved and the system retains sensitivity to the impact direction. In contrast, fluorine termination breaks the hexagonal symmetry, as observed for \ce{MoS2F}, and the directional dependence is largely lost. Substitution of \ce{W} for \ce{Mo} does not lead to significant differences in the threshold energy within the uncertainty of $\pm$3 eV.}

\textcolor{black}{A similar argument is expected to hold for most functionalizing atoms, except for very light species such as hydrogen. Because hydrogen is much lighter, it is less effective at slowing down incoming Ar atoms, so the resulting collision pathways are likely to differ from those involving O or F. In addition, hydrogen adsorption is more complex, likely due to its small size. Atomic hydrogen can occupy several competing adsorption sites, including an interstitial site in the plane of the metal (Mo) atom inside the hexagonal lattice \cite{KeongKoh2012,khalid2024Role}. Consequently, more detailed studies are required to quantitatively assess sputtering in H functionalized TMDs such as \ce{MoS2H}.}

More specifically, \ce{MoS2} and \ce{MoS2O} exhibit ranges of possible threshold energies: $32 - 24$ eV and $14 - 7.5$ eV, respectively, which depend on the impact angle (Figure \ref{fig:F4}). However, achieving the threshold energies of $\sim (32 \pm 1)$ eV for \ce{MoS2} and $\sim (14 \pm 1)$ eV for \ce{MoS2O} under strictly normal \ce{Ar} incidence is likely challenging under realistic plasma conditions (Figure \ref{fig:F5}b, $\thtInherent > 7^{\circ}$), where projectiles exhibit an angular spread. In practice, the effective threshold values will be lower because temperature effects disrupt perfectly orthogonal impacts. First, plasma ions have a transverse temperature of at least $\sim 0.1$ eV, which creates a spread in impact angles $\sim 5^{\circ}$ even under low-pressure conditions \cite{Woodworth1996}. Secondly, even if some ions impinge on the TMD surface in a truly orthogonal direction and hit O atoms, the O atoms deflect with an angular spread $\theta_T$ (Figure \ref{fig:F5}b, $\thtInherent = 0$). This spread, $\theta_T$, is determined by the TMD temperature, because thermal fluctuations of the target O atoms grow with increasing temperature, roughly following a $\sim \sqrt{T}$ scaling. Therefore, the thresholds observed in experiments are more likely to be closer to $\approx (26 \pm 2)$ eV for \ce{MoS2} and $\sim (9 \pm 1)$ eV for \ce{MoS2O}. By tightly restricting the \ce{Ar} incidence angle to within a spread of $< 5^{\circ}$ and minimizing thermal fluctuations by cooling the TMD to cryogenic temperatures of approximately $(-200$ to $-50{})^{\circ}\mathrm{C}$, it may be possible to increase the threshold closer to $\sim (31 \pm 2)$ eV for \ce{MoS2} and $\sim (13 \pm 1)$ eV for \ce{MoS2O}. To observe the proposed temperature sensitivity, one can also use narrow beams with a very small angular divergence.

One should also keep in mind the limitations of the reported findings. First, the precise numerical values of the reported energy thresholds can depend on the specific details employed in the DFT calculations. However, the key conclusions that sputtering is substantially enhanced by oxygen/fluorine functionalization (and also by increased impact angle) follow from straightforward physical arguments, indicating that they should remain valid regardless of the specific DFT implementation. Second, the proposed sputtering mechanism where products such as \ce{SO2} and \ce{SF_n} form and then desorb requires a high surface coverage of oxygen and fluorine ($\sim 10\%$ or higher). Such coverages are reasonable, given the strong adsorption of atomic oxygen on \ce{MoS2} \cite{KC2015} and other TMDs such as \ce{WS2} \cite{Rawat2024}. However, substitutions of S atoms with O atoms in the TMD itself are also favorable in certain cases \cite{Kang2014}, so kinetic tuning may be required to reach high adsorption coverage without significant TMD structure changes. Achieving the necessary surface coverage can be accomplished with fast gas valves and low energy plasma sources \cite{Zhao2021}. \textcolor{black}{Cleaning the functionalizing O/F atoms after achieving the desired processing effect also requires additional steps such as high-vacuum annealing \cite{Yang2015}. The removal of residual functionalizing atoms is essential to preserve downstream device quality, motivating further systematic studies of post-processing cleaning approaches. Additionally, the stability of the system in the presence of chemically reactive gases, such as \ce{SF6} used during functionalization, warrants further investigation. Notably, a similarly reactive gas, \ce{MoF6}, has been employed in atomic-layer etching \cite{Soares2023} and was observed to induce damage only when alternated with another reactant, such as \ce{H2O}.} Furthermore, at low adsorption coverage, different mechanisms are expected, since the formation of \ce{SO2} and \ce{SF_n} under these conditions is unlikely. Similarly, defects from growth stage or grain boundaries may favor different sputtering pathways with low energy threshold, but this goes beyond the scope of this work. Finally, a realistic ion angular spread of about $5$–$10^{\circ}$ \cite{Woodworth1996} would decrease the temperature sensitivity, as shown in Figure \ref{fig:F5}b. More MD simulations may help to quantify such effects.


\begin{acknowledgement}

This material is based upon work supported by the U.S. Department of Energy, Office of Science, Fusion Energy Sciences and Basic Energy Sciences, as part of the Extreme Lithography \& Materials Innovation Center (ELMIC), a Microelectronics Science Research Center (MSRC), under contract No. DEAC02-09CH11466. The simulations were carried out at the National Energy Research Scientific Computing Center (NERSC), a U.S. Department of Energy Office of Science User Facility located at Lawrence Berkeley National Laboratory, operated under Contract No. DE-AC02-05CH11231 using NERSC award BES-ERCAP36136 and the Stellar, Della, and Tiger clusters at Princeton University. We would like to thank John Mark Martinez, Jack Draney, Louis Hoffenberg, and David Graves for fruitful discussions.

\end{acknowledgement}

\begin{suppinfo}

\section{SI files descriptions}

V1.mp4: Impact of a single 10eV Ar at a MoS2O single-layer TMD equilibrated at 350K at an angle. It results in formation and sputtering of SO2, demonstrating the ''chemically enhanced physical sputtering'' mechanism.

V2.mp4: Impact of a single 34eV Ar at a MoS2 single-layer TMD equilibrated at 350K normally. It results in ``reflection'' of Ar from the Mo layer, followed by formation and sputtering of S3. It demonstrates the dominant sputtering pathway at low-energy sputtering.

\section{Sputtering yield at near-threshold energies}

Here we provide a simple estimate of yield at projectile energies slightly above the sputtering lower energy threshold $\EsputtS$. More precisely, we provide a lower bound on the yield, meaning we predict at least this much yield, though it is possibly higher if other sputtering pathways activate at higher energies. The main task is to connect the sputtering probability from the most susceptible hit-point $\Psputt(\EAr | \text{optimal-hit})$ (Figure \ref{fig:F2}) to an experimentally measurable yield, which comes from uniform impacts of all hit-points. The key assumption we make for this estimate is that hit-points closer than in-plane thermal fluctuations of the top-layer atoms behave similarly. This allows us to estimate that if sputtering from the most-susceptible hit-point becomes possible at $\EAr > \EsputtS$, then sputtering from points within $\sim \langle \sigma_{xy} \rangle$ from the most susceptible point also becomes possible. For the order-of-magnitude estimate, we also assume the sputtering probability from these points $P_{S}(x,y) = \Psputt(\EAr | \text{optimal-hit})$ does not depend on the hit-point within this small circle. For a given elementary triangle, the 1/6 fraction of the points within $\langle \sigma_{xy} \rangle$ lies inside the triangle. This gives the lower bound on the yield

\begin{equation}
    Y_S (\EAr) \geq \Psputt(\EAr | \text{optimal-hit}) \frac{\pi \langle \sigma_{xy} \rangle^2 / 6}{S_{\text{triangle}}}
\end{equation}

Approximate equality is realized when no additional sputtering pathways, except for one found at $\EAr = \EsputtS$ activate at $\EAr > \EsputtS$. If such new pathways activate, yield can be much higher. 

We aim at a lower bound for yield, so we use a conservative (low) $\langle \sigma_{xy} \rangle \sim 0.1$\AA{} (Figure \ref{fig:F4S1}). We use $a_{\text{lattice}} = 3.15$\AA{} (Figure \ref{fig:F2S2}) for $S_{\text{triangle}} = a_{\text{lattice}}^2 / 4 \sqrt{3}$ and get $Y_S (\EAr) > 0.0036 \cdot \Psputt(\EAr | \text{optimal-hit})$. Given $|\Psputt| \sim 1$, the order of magnitude $Y_S \sim 10^{-2.5}$ is consistent with the yield reported near the lower sputtering energy threshold for \ce{MoS2} in MD simulations \cite{Kretschmer2018}.

As addressed in Figure \ref{fig:F2S4}, precise quantification of yield near the sputtering threshold depends on exact definitions of sputtering vs assisted desorption. Additionally, yield per impact depends on the degree of damage already present near the hit-point, which makes impacts history-dependent and forces simulating many consecutive events for the same material. Finally, systematic quantification of yield requires sampling of all hit-points in the elementary triangle (Figure \ref{fig:F1}d). Such calculations require orders of magnitude more sampling than reported here, which is why they are often conducted via force-field MD \cite{Kretschmer2018}.

The same logic also allows us to estimate how to compare our sampling $N_{\text{hits}}$ per a fixed hit-point to the usual sampling per elementary triangle $N_{\text{sample}} \sim N_{\text{hits}} \frac{S_{\text{triangle}}}{\pi \langle \sigma_{xy} \rangle^2} \approx 270 \cdot N_{\text{hits}}$. Thus, given the minimal sampling $N_{\text{hits,min}} = 14$ used in this work, this translates into $N_{\text{sample,min}} \approx 4000$.

\section{Surface coverage fraction effects}

The sputtering through the formation of (meta)stable species such as \ce{SO2} relies on a high surface-coverage assumption because forming such products requires having several functionalizing atoms near each other. Lower surface coverage will decrease the probability of the proposed sputtering mechanisms. The fluorine case is more complicated and most likely requires AIMD studies to quantify coverage fraction effects. The \ce{MoS2O} case, however, has a simpler, cleaner sputtering mechanism, which allows us to estimate the oxygen surface coverage fraction necessary to make the \ce{SO2} sputtering mechanisms dominant. At a coverage fraction $c$, each \ce{SO} pair has $N=3$ neighboring sites that can host O atoms. The probability that $n$ of them are occupied is $C_n^N c^n (1-c)^{N-n}$. Only reflections of the impacted O into a neighboring \ce{SO} activates the proposed sputtering pathway, so the probability to activate it decreases by $n/N$ compared to full coverage. Thus, the probability to result in sputtering given an impact at the most susceptible hit-point at a random thermal fluctuation is

\begin{equation}
    \Psputt(\dots, c < 1) \approx \Psputt(\dots, c = 1) \cdot \sum_{n=1}^N \frac{n}{N} C_n^N c^n (1-c)^{N-n} = ... = \Psputt(\dots, c = 1) \cdot c
\end{equation}

Interestingly, this result does not depend on $N$. Other summation weighting may result in a different coefficient in front of $c$, which may depend on $N$. But the expression always will be $\alpha c + \mathcal{O}(c^2)$ with $\alpha\sim 1$. This tells us that if we want the yield $Y_S$ of the given sputtering pathways to decrease no more than by $\gamma = 0.1$ times due to decreased surface coverage, then from $\Psputt(\dots, c < 1) / \Psputt(\dots, c = 1) > \gamma = 0.1$ the surface coverage should be $c > \gamma = 0.1$.

\section{Sputtering mechanism differences for \ce{MoS2O} vs \ce{MoS2F}}

We hypothesize that forming \ce{SO2} is the key step for \ce{MoS2O} sputtering (Figure \ref{fig:F3}), which will inevitably lead to thermal desorption of \ce{SO2}. This means that only a single S-O bond needs to be broken by \ce{Ar} to enable such a sputtering mechanism. Forming \ce{SF_n} products, on the other hand, may not be sufficient for their subsequent thermal desorption because they are much more reactive. Thus, weakening of the Mo-S bonds by F atoms (due to high electronegativity of fluorine) may be a necessary part of the decrease in the sputtering energy threshold.

A chemical reason for symmetry breaking in \ce{MoS2F}, but not in \ce{MoS2O} can be impacted by the electronegativity difference between F and O. The more electronegative F can pull more electron density from S. The S-F bond is also longer than S-O ($d_{bond} \approx 1.9$ \AA{} vs 1.4 \AA{}, see Figure \ref{fig:F1S1}a). This creates a stronger dipole $\sim d_{bond} q_{bond}$ in each S-F pair. Such dipole formation was reported for \ce{MoS2O} \cite{Yan2025}. All dipoles are aligned, making them mutually repulsive, and this repulsion is stronger for S-F dipoles. Thus, this may impose sufficient stress on \ce{MoS2} to break its symmetry. 

Additionally, adding an F atom to the \ce{MoS2} unit cell adds odd number of electrons, while adding an O adds an even number of electrons. There is a well-known ``Peierls instability'' that was first explained in 1D \cite{Kagoshima1981, Peierls2001}, but later ``Peierls distortions'' were found in 2D and 3D \cite{Burdett1983, You2021}. All these effects are driven by the fact that unit cells with odd number of electrons can undergo ``spontaneous'' symmetry breaking that splits the degeneracy in electron bands near the edge of the Brillouin zone and thus the valence band decreases its average energy. We believe that distortions of F atoms from positions directly above their host S atoms is a case of such ``Peierls distortions''.

We note that typical sputtering events reported here occur on the timescale $< 1$ ps, as shown in Figures \ref{fig:F3} and \ref{fig:F3S1}. This does not mean sputtering is impossible at lower energies; for example, even thermal sputtering on a timescale of months is feasible\cite{Pet2018}. We also observed a multi-step S-vacancy creation process on timescales as long as $\sim$ 20 ps (shown on Figure \ref{fig:F2S4}). However, such events exhibit a clear separation of timescales between the product formation and product removal. An \ce{Ar} impact that is not sufficient for fast ejection of products can still catalyze product formation, with subsequent thermal desorption occurring on a much longer timescale. Since most of the immediate sputtering events we observed occurred within $<$ 1 ps, we chose 2 ps as the timescale to separate "immediate" sputtering from thermal desorption of products. Thus, we simulate for 2 ps after the impact and conclude that no sputtering has occurred if the products do not detach from the material within this time. 

\section{Angular sensitivity $\Leftrightarrow$ ground-state order}

Fluorinated \ce{MoS2}, which is less ordered than \ce{MoS2O} even before the impact, shows no systematic variation with increasing impact angle. This indicates that once the symmetry is disrupted, further changes in the direction of incoming projectiles have little additional effect on the sputtering behavior. This is because the main effect of a slight tilt of the impact from the surface normal is to allow the impacted O/F to pass by the S atom directly below it, thereby enabling the O/F to break its bond with the host S atom and potentially initiate the formation of sputtering products. Such tilting is necessary for \ce{MoS2O} where O atoms sit directly on top of S atoms, whereas F atoms are already slightly shifted from their underlying S atoms, so small additional tilts do not produce a significant change.

The in-plane angle $\varphi$ is also shown to be important for \ce{MoS2O} (Figure \ref{fig:F4S2}). We found in-plane angles that maximize sputtering to be $\phiOpt(O) = 30^{\circ}$ and $\phiOpt(S) = 0^{\circ}$. Fluorine did not show sensitivity to $\varphi$. This is consistent with proposed sputtering mechanisms: Pristine \ce{MoS2} is sputtered by physically breaking the bonds of a single S atom, which requires that other in-plane atoms do not obstruct the S atom as it gains maximum displacement, while still preventing it from penetrating too deeply into the material in order to minimize intermediate collisions. This condition is achieved at $\varphi = 0^{\circ}$ (see Figure \ref{fig:F1}d) where the impacted S atom has the largest available in-plane space in the reflection direction. The hit-point at the center of the hexagon (which was found to be optimal for \ce{MoS2}) leads to the projectile energy being distributed approximately equally among three Mo atoms, such that no single Mo atom deviates too far from its equilibrium position and collides with its neighbors. The temporary storage and redistribution of projectile energy in this case is more complex and likely involves several many-body effects. The mechanism for \ce{MoS2O}, on the other hand, involves the formation of \ce{SO2}, which subsequently escapes. Impacting an O atom at $\varphi=30^{\circ}$ directs it toward its nearest S neighbor that has another O atom attached (see Figure \ref{fig:F1}d). Thus, \ce{SO2} formation is expected to be maximized at $\varphi=30^{\circ}$, with the hit-point directly on the oxygen. We also quantify the $\varphi$ dependence (Figure \ref{fig:F4S2}b) for a fixed $\theta = 10^{\circ}$ bit enough to see benefit from tilting $\theta$. Approximately 25\% of the $\varphi$ range ($\sim [25; 40]^{\circ}$ out of $[0;60]^{\circ}$) shows optimal energy threshold values within errorbars. Threshold for all other $\varphi$ values can be obtained by symmetry (Figure \ref{fig:F4S2}c). Blue dashed lines on Figure \ref{fig:F4S2}c show how 6 energy threshold minima match with the 6 directions at 6 neighboring oxygen atoms (given the hit-point is at an oxygen atom). This further confirms out interpretation. Finally, the insensitivity of \ce{MoS2F} to $\varphi$ again highlights how structural disorder suppresses its directional response.

\section{Finding the most damage-susceptible hit point}

An analysis similar to Figure \ref{fig:F2S5} is also performed for \ce{MoS2F}. It shows that the site above sulfur is among the most susceptible. Different F atoms are displaced by different amounts from being directly above their S atoms, which means that a head-on collision requires a hit-point slightly offset from the directly above-sulfur position. We speculate that the effect of this shift is negligible, because the directional insensitivity of \ce{MoS2F} arises not only from the random strong displacements of the equilibrium positions of the directly impacted F atoms, but also from the surrounding F atoms. These surrounding F atoms will still not be sufficiently ordered even if the initial Ar-F impact is head-on. Analogous tests for \ce{MoS2} find the most susceptible point to be at the center of the 3Mo-3S hexagon (the \ce{Ar} position in Figure \ref{fig:F3S1} at t=0). We speculate that this is because it is the most effective hit-point for the sputtering mechanism shown in \ref{fig:F3S1}. This mechanism is likely the most energy-efficient, as the reversal of the projectile velocity occurs via "reflection" from the Mo layer through approximately equal displacement of three Mo atoms. As a result, the maximum Mo displacement is minimized, reducing Mo-Mo collisions with neighboring atoms and thereby minimizing energy losses.

\section{Simple theory for \ce{MoS2O} $\Esputt(T)$}

The qualitative difference in the angular dependence $\Esputt(\theta)$ between \ce{MoS2O} and \ce{MoS2F} has important implications for their temperature dependence since thermal fluctuations lead to a finite spread in reflection angles even for orthogonal impacts. This angular spread is non-negligible because the magnitude of thermal fluctuations of O and F atoms in the material plane is comparable to their hard-core repulsion interaction distance with \ce{Ar} (see Figure \ref{fig:F4S3}). This means that even orthogonal impacts will not always result in strictly vertical momentum transfer. Instead, thermal fluctuations of O/F atoms make the collision non-head-on, thus introducing a distribution of reflection angles (see SI and Figure \ref{fig:F4S3}): 

\begin{equation} \label{eq:thtT_exct}
\theta_T(\EArPerp) \approx \arcsin \left( \frac{\sigma_{xy, X}(T) s_N}{rArX(\EArPerp, \theta_T(\EArPerp))} \right)
\end{equation}

where $\sigma_{xy, X}(T)$ represents the magnitude of thermal fluctuations of X=\{O,F\} in the plane of \ce{MoS2}, (see Fig \ref{fig:F4S1} for its temperature dependence). The scale factor $s_N = \chi_1.PPF(1 - 1/(N+1))$ comes from our definition of the sputtering threshold. We have a finite sample of $N$ impact trials, and we register sputtering if at least one trial results in sputtering. This means that the maximum fluctuation of magnitude $\sigma$ over all $N$ trials has the probability of exceeding $\sigma_N$ equal to the probability of a single fluctuation exceeding $\sigma$. Therefore, we use $\sigma_N = \sigma s_N$ as a characteristic fluctuation size instead of just $\sigma$. In a normal case, $\sigma_{xy}$ would be distributed $\sim \chi_2 = \sqrt{\chi^2_2}$ due to $\sqrt{\sigma_x^2 + \sigma_y^2}$. However, only a narrow range of $\varphi_{opt} \pm \delta\varphi$ angles is susceptible to damage near the sputtering energy threshold, with $\delta\varphi \sim 8^{\circ}$ (Figure \ref{fig:F4S2}b), so only deviations along a single axis $\varphi \sim \varphi_{opt}$ are relevant, and the distribution reduces to $\chi_1$ in our case. Since we require the probability of observing a deviation larger than $\sigma_N$ at least once in $N$ trials to be significant, we set $\chi_1.CDF(\sigma_N / \sigma) = 1 - 1/(N+1)$, where $N+1$ instead of $N$ is used to remove singularity at $N+1$. This means that a single fluctuation has a probability $1/(N+1)$ of exceeding $\sigma_N$. Thus $\sigma_N = \sigma \cdot \chi_1.CDF^{-1}(1 - 1/(N+1)) = \sigma \cdot \chi_1.PPF(1 - 1/(N+1))$. For large $N$, this quantity has a weak dependence of $\approx \sqrt{2 \ln(N+1)}$, therefore the exact number of trials is not substantially important.

The term $rArX$ denotes the Ar-X interatomic distance, defined as the sum of hard sphere radii that approximate the interacting atoms during the collision. It is determined by how closely the head-on component of the collision energy, $\EArPerp \cos^2(\theta)$ brings the atoms together. Thus, it depends on the impact energy and collision geometry as described below (see SI for details):

\begin{equation} \label{eq:U_rArX}
\UArX(\rArX) = \frac{\EArPerp \cos^2{\theta}}{1 + \mX / \mAr}
\end{equation}

where $\UArX$ is the two-atom potential energy surface (PES) between \ce{Ar} and X. We take its inverse function $\UinvArX$ such that $r = \UinvArX(E)$ is the interatomic distance at which $\UArX(r) = E$. The factor $1 + \mX / \mAr$ comes from the center-of-mass frame treatment of collision.

For typical relevant $(\EArPerp, T)$ ranges, the in-plane thermal fluctuations are smaller than the collisional radii of the atoms, which results in a narrow spread of the reflection angle $\theta_T$. More precisely, $\sigma_{xy,X}(T) \ll \rArX(\EArPerp, \theta_T(\EArPerp))$, which leads to $\theta_T \ll 1$. Therefore, we can write $rArX(\EArPerp, \theta_T(\EArPerp)) \approx \rArX(\EArPerp, 0) = \rArX(\EArPerp)$. We can also assume a harmonic potential well for small $\sigma_{xy,X}(T)$, which leads to $\sigma(T) = \sigma_0 \sqrt{T/T_0} \propto \sqrt{T}$ where $\sigma_0 = \sigma_{xy,X}(T_0)$. The scaling $\sim \sqrt{T}$ is confirmed in Figure \ref{fig:F4S1}. Combining this, we can write an approximate explicit expression 

\begin{equation} \label{eq:thtT_approx}
\theta_T(\EArPerp) \approx \arcsin \left( \frac{\sigma_{0} s_N}{\rArX(\EArPerp)} \sqrt{\frac{T}{T_0}} \right) \approx \frac{\sigma_{0} s_N}{\rArX(\EArPerp)} \sqrt{\frac{T}{T_0}}
\end{equation}

Thus, an O or F atom fluctuating at temperature $T$ and impacted by an \ce{Ar} atom with orthogonal energy $\EArPerp$ can acquire velocities distributed within an angular range of $\theta \approx 0 \pm \theta_T(\EArPerp)$.

On the other hand, it is possible to isolate the effect of impact non-normality from thermal noise by bombarding surfaces equilibrated at temperatures low enough that thermal fluctuations are negligible, meaning $\sigma_{xy}(T) \ll \dAtomX = \rArX(\EsputtMin)$. The results are shown in Figure \ref{fig:F4S0}a. No significant angle dependence is observed for \ce{MoS2F}, as expected from its irregular ground state structure. However, \ce{MoS2O} exhibits a non-linear decrease in $\Esputt(\theta)$, dropping by $\sim 40-50 \%$ at $\theta \sim 10-15^{\circ}$ before flattening. Although not shown in the plot, it starts rising at some point $\thtOpt$ between $30^{\circ}$ and $45^{\circ}$. The curve $\EArX(\theta < \thtOpt)$ therefore represents the energy threshold that \ce{Ar} must have in a head-on, non-normal collision with X to induce ejection of an S atom.

Thus, eq.\eqref{eq:thtT_exct} gives the range of reflection angles that are thermally accessible (purple shade on Figure \ref{fig:F4S0}b), while the inverted curve $\EArX(\theta)$ from Figure \ref{fig:F4}a provides the minimum reflection angles $\thtArX(\Eheadon)$ required for sputtering (blue shade on Figure \ref{fig:F4S0}b). As the angular spread increases with $T$, once angles $\theta > \thtArX(E_{1})$ become thermally accessible at a certain temperature $T_1$, sputtering with $\Eheadon = E_1$ becomes frequent at $T > T_1$. See the SI text for the derivation of the purple curve in Figure \ref{fig:F4S0}b. Solving eq.\eqref{eq:mainTh} for X=Oxygen numerically gives the red dashed line in Figure \ref{fig:F5}a.

\section{2-body collision model}

To translate the head-on, non-normal energy threshold $\EArX(\theta)$ into a threshold for a non-head-on but normal impact, $\EArPerp(\theta)$, we recall that only a factor of $\cos^2(\theta)$ of the total energy participates in the collision when the reflection angle is $\theta$. This yields $\EArPerp(\theta) \cos^2(\theta) = \EArX(\theta)$. This results in a system of two equations with two unknowns $(\theta_T, \EArPerp)$:

\begin{equation} \label{eq:mainTh}
\left\{
\begin{aligned}
    & \UArX \left( \frac{\sigma_{xy}(T)}{\sin(\theta_T)} \right) = \frac{\EArPerp \cos^2(\theta_T)}{1 + \mX / \mAr} \Rightarrow \theta_T(\EArPerp) \\
    & \EArPerp = \frac{\EArX(\theta_T(\EArPerp))}{\cos^2(\theta_T(\EArPerp))} \Rightarrow \EArPerp(T) \\
\end{aligned}
\right.
\end{equation}

where the first equation describes the spread of reflection angles of the impacted O/F atoms and the second equation relates the normal, non-head-on impact energy (which is often controlled experimentally) to the non-normal head-on energy (which is easier to simulate and is temperature independent to a good approximation).

The two equations correspond to two physical conditions required for sputtering in the proposed mechanism: thermal fluctuations of the impacted O/F should be wide enough to allow the reflection angle spread $\theta_T$ that includes angles sufficient for sputtering according to $\EArX(\theta)$. This breaks down into 2 conditions:

\begin{enumerate}
    \item At a given temperature $T$, normal \ce{Ar} impacts induce a spread of reflection angles $\theta < \theta_T$. The purple region in Figure \ref{fig:F4S0}b corresponds to such a region for $T = 300$K. There is also a slight dependence on $\EArPerp$, because higher $\EArPerp$ result in "smaller" repulsive sizes of atoms.
    \item At each angle $\theta$, there is a minimal energy $\EArX(\theta)$ that an \ce{Ar} must have in a head-on collision at angle $\theta$ to sputter an S atom. This corresponds to the blue region on \ref{fig:F4S0}b
\end{enumerate}

Combining the two, the blue region in Figure \ref{fig:F4S0}b represents pairs (head-on energy; angle) = $(\EArX, \theta)$ that would result in sputtering, while the purple region represents the accessible spread of reflection angles that are realized during an impact. Therefore, sputtering happens for those $(\EArX, \theta)$ values over which the two regions overlap, meaning that thermal fluctuations make the reflection angle spread wide enough to reach angles $\theta$ that result in sputtering at a given energy.

Finally, the inherent $\theta$-spread of the projectiles is incorporated by using $\sqrt{\theta_T(\EArPerp)^2 + \thtInherent^2}$ as the new $\theta_T$. Such simple quadratic addition of fluctuations is used because of two reasons. First, the angles are small and therefore add approximately linearly. Additionally, the inherent and thermal $\theta$ fluctuations are statistically independent, so their variances add linearly.

\subsection{Derivation details}

Derivation of eq.\eqref{eq:U_rArX} - \eqref{eq:mainTh}: Referring to Figure \ref{fig:F4S3}, we first transform to the center-of-mass(COM) frame as shown below:

\begin{equation}
    \mO \vCoM = \mAr(\vAr - \vCoM) \hspace{10pt} \Rightarrow \hspace{10pt} \vCoM = \frac{\mAr \vAr}{\mAr + \mO}
\end{equation}

 We then decompose the collision into a head-on collision along the line connecting the centers of \ce{Ar} and O (thick dashed line in Figure \ref{fig:F4S3}) and the orthogonal direction, which does not experience any changes after the collision. Along the head-on direction, we can write energy conservation to find the effective hard-sphere distance between \ce{Ar} and O at their closest approach $\rArO$:

\begin{equation}
    \frac{\mO (\vCoM \cos(\theta))^2}{2} + \frac{\mAr ((\vAr - \vCoM) \cos(\theta))^2}{2} = \Ecollision = \UArO(\rArO)
\end{equation}

Relativistic corrections are neglected because $\vAr / c \sim 10^{-4}$ at $\EArPerp \sim 10$ eV. Thus, the incoming \ce{Ar} energy is $\EArPerp = \mAr \vAr^2/2$. Plugging everything in, we get

\begin{equation}
    \Ecollision = \frac{\EArPerp \cos^2(\theta)}{1 + \mO / \mAr}
\end{equation}

which is what appears in eq.\eqref{eq:U_rArX}.

The derivation of eq.\eqref{eq:thtT_exct} follows directly from Figure \ref{fig:F4S3}: The triangle on points (Ar position; O current position; O equilibrium position) has the angle $\theta$ and $\pi/2$, and sides $\sigma_{xy,O}$ and $\rArO$, thus $\sin(\theta) = \sigma_{xy,O} / \rArO$, which is equivalent to eq.\eqref{eq:thtT_exct}.

\subsection{Numerical solution for $\EArPerp(T)$}

Equations \eqref{eq:mainTh} can be rewritten in a less intuitive form, which nevertheless simplifies mathematical analysis 

\begin{equation}
\left\{
\begin{aligned}
    & \UArX \left( \frac{\sigma_{xy,X}(T)}{\sin(\theta_0)} \right) = \frac{\EArX(\theta_0)}{1 + \mX / \mAr} \hspace{10pt} \Rightarrow \hspace{10pt} \theta_0 (\sigma_{xy,X}(T)) \\
    & \EArPerp(T) = \EArPerp(\theta_0 (\sigma_{xy,X}(T))) = \frac{\EArX(\theta_0)}{\cos^2(\theta_0)} \\
\end{aligned}
\right.
\end{equation}

where $X$ is the impacted atom type, which can be O/F/S, and $\theta_0$ is different from $\theta_T$ above and is just a function of $\sigma$. 

A sufficient condition for solution uniqueness is $\theta_0 < \thtOpt$ where $\thtOpt(X) = \texttt{argmin}_{\theta}(\EArX(\theta))$. Oxygen is shown to satisfy at least $\thtOpt(O) > 30^{\circ}$. This gives the maximum applicable temperature condition 

\begin{equation}
T < \Tmax(X) = \sigma_{xy,X}^{-1} \left[ \frac{\sin(\thtOpt(X)) \UArX^{-1}(\EArX(\thtOpt(X)))}{1 + \mX / \mAr} \right]
\end{equation}

where $\sigma_{xy,X}^{-1}(\sigma)$ is the inverse function of $\sigma_{xy,X}(T)$ such that $\sigma_{xy,X}^{-1}(\sigma_{xy,X}(T)) = T$.

We only need to consider $U(r < \rWell)$, where $\rWell = \texttt{argmin}_{r}[U(r)]$, because this is the range that contains the hard-sphere distance $\rHardsphere$ that approximates the collision. The range $r < \rWell$ still covers the full range or energy values $[0; \infty]$, so both $\sigma_{xy}(T)$ and $U(r < \rWell)$ are always invertible. This is shown by shading the $r > \rWell$ on Figure \ref{fig:F4S3}b. 

For oxygen, we can estimate $\Tmax(\thtOpt(O)) > \Tmax(30^{\circ}) \approx 4900$ K, which is beyond \ce{MoS2} melting point of $\Tmelt \sim 2650$ K \cite{MerckIndex}. Therefore, the theory is formally valid for all physically relevant temperatures. In practice, the breakdown is expected to occur earlier, when the harmonic-well approximation $\sigma_{xy}(T) \sim \sqrt{T}$, used here to extrapolate $\sigma_{xy}(T)$ in figure \ref{fig:F4S1}, begins to fail. A numerical evaluation of $\sigma_{xy}(T)$, would allow the applicable temperature range to be extended close to $\Tmelt$.

\subsection{High and low $T$ regimes of $\Esputt(T)$}

The linear regime for low $T$ comes from the fact that $\EArX(\theta)$ in Figure \ref{fig:F4} is quadratic at the origin $\theta = 0$, and $\cos(\theta)$ is also quadratic at $\theta = 0$. Thus, for temperatures $T$ such that $\theta_T \ll 1$ we can expand $\EArPerp(T) \approx E_0(1 - \alpha \theta_T^2)$. Next, we recall from eq.\eqref{eq:thtT_approx} that for low $T$, $\theta_T \approx \sigma_0/\rArX(\EArPerp) \sqrt{T/T_0}$. The dependence of $\rArX(\EArPerp)$ turns out to be weak for $\EArPerp \sim 10$eV, allowing us to approximate $\rArX(\EArPerp) \approx \dAtomX = \mathrm{const}$ as shown below: 

\begin{equation} \label{eq:Ear_approx}
    \EArPerp(T) \approx E_0 \left( 1 - \alpha \left( \frac{\sigma_0}{\dAtomX} \right)^2 \frac{T}{T_0} \right) = E_0 - b T
\end{equation}

The plateau regime is more incidental. It occurs when the decrease in $\EArX(\theta_T)$ is compensated by the decrease in $\cos^2(\theta_T)$. It is wide compared to the linear region because $\sigma_{xy}(T) \propto \sqrt{T}$ changes more slowly at higher $T$, and $\EArX(\theta)$ likewise varies more slowly at higher $\theta$. The transition from the linear regime happens at $T_{switch}$ when $\EArX''(\theta)$ deviates from its initial parabolic shape, meaning the curvature $\EArX''(\theta_{T_{switch}})$ becomes significantly different from $\EArX''(0)$.

The theoretical prediction in Figure \ref{fig:F5}a is not followed exactly within the error-bars, which can be due to several approximations. For very small $\theta$, the two-body approximation may break down, since the impacted O/F atom is pushed almost directly into the S atom below it, making the process at least a three-body problem. Too wide angles $\theta \sim 1$ may start to break the hard-sphere collision model because higher $\theta$ leads to longer periods before the main collision when the \ce{Ar} already feels the material below, which may alter its course in a more complicated way.

\section{Spin treatment}

Treating spins breaks down into two separate questions that are distinguished by their timescales: the appropriate equilibrium initial states and the appropriate dynamics during and after the collision. First, we explain why singlet initial states were chosen, and then why we decided that fixed-spin singlet dynamics was the optimal choice between computational cost and accuracy for dynamics.

First, we note that we used CP2K \cite{Khne2020, Hartwigsen1998, Frigo2005, Krack2005, VandeVondele2007, Grimme2011, Bortnik2014, Heinecke2016, Schtt2016, Goerigk2017} to obtain results presented in this work, and we are only aware of global spin fixation within this framework. This can lead to unpaired electrons localizing on parts of the system where they are not desired, such as the impacting argon atom. Impacts with metastable projectiles can happen in reality, but are beyond the scope of this work. Simulating \ce{Ar} impacts in a fixed triplet state lowers the sputtering thresholds relative to singlet results. However, the ground states of 2H phases of \ce{MoS2}, \ce{MoS2O} and \ce{MoS2F} are singlets (meaning these materials are not ferromagnetic). Thus, initializing the system in a triplet state puts it in an excited state configuration. We estimate it to be $\sim$ 1.8 eV above the ground state for a 4x4 single-layer \ce{MoS2} + \ce{Ar} system. 

Oxygen is known to exhibit non-trivial spin effects. For example, its molecular form \ce{O2} is known to have a triplet ground state. Figure \ref{fig:F1S1}b red lines show PES for an O atom at varying distances from \ce{MoS2}. It is significantly different for singlet (solid) and triplet (dashed) spin states of the systems. All simulations reported in the main text were performed in a fixed-spin singlet state, which may pose a question about their validity, since potential oxygen dissociation from \ce{MoS2O} can play a significant role in the processes investigated in this work. However, we provide the following qualitative arguments for why simulating impacts in a fixed singlet state represents an optimal choice between computational cost and accuracy:

\begin{enumerate}
    \item The main ejection product for \ce{MoS2O} is \ce{SO2}, which has a singlet ground state. We simulated fixed-triplet \ce{Ar} bombardment of \ce{MoS2O} and the dominant product was still $\ce{SO2}$. The sputtering threshold in fixed-triplet simulations decreased by $(2 \pm 0.5)$ eV. This is consistent with our observation that the total system energy in the triplet state is 1.8 eV higher than in the singlet state, corresponding to an "unpairing excitation" of 1.8 eV.
    
        \item Oxygen singlet is a ground state when attached to \ce{MoS2} (Figure \ref{fig:F1S1}b) and it follows the singlet PES at least until $r_{S-O} \approx 2.15$ \AA{}. Triplet becomes lower at larger $r$, which might be relevant for atomic O sputtering, for example during oxygen cleaning from the TMD after processing. However, we do not focus on O sputtering in this work.
    
    \item While the triplet becomes lower than singlet for O for $r > 2.15$ \AA{}, we believe that fix-spin dynamics is appropriate even for the impacts simulated here, because of the spin-relaxation dynamics of oxygen:
    \begin{enumerate}
        \item While estimating de-excitation time of a specific atomic O near a TMD has not been studied previously, we can try to get insight from molecular oxygen. Given a singlet state, \ce{O2} requires $\approx 20 \mu s$ to de-excite to a triplet state \cite{Farmilo1973, Petrov2015}. This is about $\sim 7$ orders of magnitude longer than the "immediate" sputtering processes (excluding long-term thermal desorption of products) considered in this work.
        \item The $20 \mu s$ estimate is under near-ambient conditions relative to $\sim 10$ eV collisions, which is a significant difference because de-excitations can only happen via relatively strong interactions coupling spin degrees of freedom to the rest of the system. Strong collisions can provide such coupling. Estimating de-excitation probabilities from collisions can be done in principle \cite{Dagdigian2015}, however, we did not perform such in-depth analysis because:
        \item Robust de-excitation to a triplet state can only occur at nuclei positions where the triplet electronic state is the ground state. This is not the case for O atoms attached to \ce{MoS2} and remains true until $r_{S-O} = 2.15$ \AA{} (Figure \ref{fig:F1S1}b). Therefore, the main Ar-O collision happens while O is in a position where its ground state is a singlet, thus it is not converted into a triplet.
        \item By the time the impacted O has moved more than 2.15 \AA{} away from the top S-layer, the impact energy has likely dissipated into many surrounding atoms, and no single collision is likely to be sufficiently energetic to provide strong spin-electronic coupling capable of inducing a spin-flip.
    \end{enumerate}

     \end{enumerate}

\section{DFT applicability}

The Born–Oppenheimer (BO) approximation assumes that electrons remain in their ground state for each nuclear configuration, which requires that nuclei move much more slowly than electrons. An order of magnitude estimate for the lower boundary of electron velocity in atoms is the atomic unit $v_e = \alpha c$, where $c$ is the speed of light and $\alpha = e^2 / \hbar c \approx 1/137$. An \ce{Ar} atom moving at $v = v_e$ has a kinetic energy of $\approx 1$ MeV, which is far greater than 10 eV. All other energies in the collision are smaller than this, so the BO approximation should hold well during the whole process.

The ground-state approximation is more complicated. A necessary condition for it is the prevalence of ionic stopping power relative to the electronic one. It was experimentally shown for ice bombardment with He+ ions that ion energies below $\sim 1-2$ keV are dominated by ionic scattering \cite{Fam2008}. One reason for this is the highly non-unitary mass ratio between electrons and ions, which suppresses fast energy transfer between ionic and electronic degrees of freedom.

However, excitations can be transferred directly into the electronic degrees of freedom of the material by several channels, such as impacts of excited atoms (\ce{Ar^*}) and/or neutralization of the projectile shortly before impact. The latter can also be viewed as a special case of de-excitation, since a charged particle (ion) near a material with much higher capacitance is in an excited state. De-excitations can occur through a number of Auger and resonant like channels \cite{Gainullin2020}. The resonant channel may not result in excitations because the electron transitions between states at the same energy level. It has been shown to dominate for Ne+ neutralization during \ce{MoS2} bombardment at $E \leq 2$ keV due to a match between the electronic band structure of \ce{MoS2} and the electronic levels of Ne+ \cite{Buitrago2024}. This is not always the case, for example $\sim 1/5$ He+ ions caused electron ejections from ice at $E_{ion} = 10$ eV \cite{Hagstrum1960} due to Auger neutralization. In this work, we do not account for excitations, which may affect quantitative results such as the absolute value of $\Esputt$. However, the qualitative relative results, such as different sputtering mechanisms and the decrease in $\Esputt$ they induce do rely on simple physical explanations rather than DFT results alone, and we therefore expect them to hold in reality.

\section{DFT and AIMD details}

We used UZH pseudo-potentials GTH-GGA \cite{Goedecker1996} and the corresponding basis sets TZV2P-MOLOPT-GGA-GTH \cite{VandeVondele2007}, as they are an update to the default CP2K GTH choice optimized for heavier elements such as transition metals and for mGGA functionals like R2SCAN, which we employed in several parts of this work. The Gaussian plane wave (GPW) method was used for propagation \cite{Hutter2013, VandeVondele2005, LIPPERT1997, VandeVondele2003, Goedecker1996}. The CUTOFF and REL-CUTOFF parameters were converged to 400 Ry and 40 Ry, respectively, corresponding to an energy error tolerance of $\sim 10^{-2}$ H following the official CP2K guidelines. Achieving a tolerance of $\sim 10^{-3}$ H would require 700 Ry and 60 Ry respectively. Given the energy scale of our processes $\sim 10$ eV with $\Delta E / E \sim 0.1$, we selected the coarser $\sim 10^{-2}$ H accuracy. A unit cell of 4x4 was used for main results. No k-point sampling was done, meaning a 1x1x1 grid and always the $\Gamma$ point. Oxygen adsorption energies computed at these parameters compared to a 5x5 unit cell with a 6x6x1 k-point grid differ by only $\sim 0.3$eV, which we consider as acceptable given the energy scale of $\sim 10$ eV of energies under study. ASPC extrapolation was used to predict the next step density \cite{Khne2007, Kolafa2003}. Quickstep EPS was set to $10^{-8}$, SCF EPS to $10^{-5}$ and orbital transformation (OT) diagonalization \cite{Weber2008} was used with a FULL-SINGLE-INVERSE preconditioner and DIIS optimizer \cite{Hamilton1986}. All considered TMDs are not metallic and/or magnetic, which allowed us to avoid smearing. A potential interaction cutoff of 10 \AA{} was applied, and a time integration step of 1 fs was used. Impact simulations were performed in N$V$E ensemble to accurately capture energy transfer between system components and dynamics. Each simulation was run for 2 ps after the impact, as justified in the main text. Equilibration was performed in the $N P_{xy} L_z T$ ensemble and propagated until lattice vectors' convergence. The initial Z-distance of \ce{Ar} atom from the top TMD layer was set to $10$ \AA{}.

The timestep of 1 fs is sufficiently small to ensure dynamically accurate trajectories of length $\leq 2$ ps considered in this work. Dynamical accuracy here means that trajectories of all atoms in the system do not deviate significantly from exact solution of the dynamical ODE despite the errors from the finite timestep. In principle, such errors accumulate and lead to exponential divergence of MD trajectories from physical trajectories \cite{Morozov2001} even in the NVE ensemble, which was used for impact simulations. At long simulation times, such divergence leads to effectively sampling the NVE ensemble instead of producing a solution to the original dynamical ODE. However, there is a ``dynamical memory time'' $t_m$ \cite{Morozov2001} before which such deviations are small and near-true dynamical trajectories are resolved. We want our collision events to be fully physical in the dynamical sense, not just in the NVE-sampling sense, because the proposed sputtering mechanisms can depend on e.g. synchronization of oscillations of several atoms, as it happens for forward-sputtering of S from pristine \MoS2. Therefore, we want the $t_m$ to be larger than at least a typical active collision stage, which was found to take $\sim 0.2$ fs. The dynamical time was estimated for the LJ system to be $\sim 4 \tau_{LJ}$, with log-dependence on the integration timestep. Making a rough approximation $\epsilon \sim 1$ eV, $\sigma \sim 1$ \AA{} and $m \sim 20$ a.u., we get $\tau_{LJ} \sim 1.4$ ps. It gives us $dt \sim 10^{-3} \tau_{LJ}$, thus $t_m(dt) \sim 5 $ ps \cite{Morozov2001}. Therefore, the 0.2 ps of the active collision, and even the whole 2 ps collision events fit into the $t_m$. Thus, we can treat the obtained AIMD trajectories as true dynamical trajectories (as opposed to just NVE sampling) and use them to suggest time-resolved sputtering mechanisms.

Figure \ref{fig:F1S1}a shows only a negligible difference between the PBE \cite{Perdew1996} and R2SCAN \cite{Furness2020} functionals for the PES, both in terms of adsorption energies and overall profiles. Dispersion corrections using rVV10 \cite{Sabatini2013} versus D3BJ \cite{Grimme2010} also show negligible differences. Therefore, we use the simplest option, PBE-D3. This choice is also commonly adopted in the literature for \ce{MoS2} + O systems \cite{Pet2018, Ma2013, Xie2013}. The R2SCAN functional combined with rVV10 dispersion was included for comparison because more sensitive properties, such as phonon spectra of \ce{MoS2} has been shown to require R2SCAN + rVV10 for accurate description \cite{Ning2022}. Our claim of a low adsorption energy barrier for atomic O and F is somewhat affected by the functional: PBE predicts a barrier closer to $0.1$ eV, while R2SCAN yields a value closer to 0.2 eV. However, both values are comparable to ambient thermal energy $k_B T \sim 0.026$ eV. Therefore, the claim of unimpeded adsorption is likely justified.

\section{Free-energy simulation setup}

We used CP2K + PLUMED (the open-source, community-developed PLUMED library \cite{PLUMED2019}, version 2.x \cite{Tribello2014} or alternatively version 1.x \cite{Bonomi2009}.) to run meta-dynamics (metaD) \cite{Barducci2008, Laio2002}. A 2x2 supercell with a single adsorbed O or F atom was employed. The distance between the adsorbing O/F atom $Z_{O/F}$ and the center of mass of the entire TMD layer $Z_{TMD}$, was chosen as the collective variable ($CV = \delta Z = Z_{TMD} - Z_{O/F}$) for metaD. We bias the Z-coordinate of the TMD layer to a fixed value with a parabolic potential $k_{TMD}(Z_{TMD} - Z_0)^2/2$ with $k_{TMD} = 2.5$ eV/\AA${}^2$. From unconstrained equilibration runs, we find that the stable bonding sites for both O and F are approximately on top of the corresponding S atoms. Therefore, we applied another parabolic bias ($k = 5 $ eV/\AA${}^2$) on $X_{O/F} - X_{S,bond}$ and $Y_{O/F} - Y_{S,bond}$ to prevent the O/F from slipping off the bonding S atom under the pressure of the metaD bias. Finally, we applied one-sided parabolic walls on $\delta Z$ at 1.5 \AA{} ($k = 3$ eV/\AA${}^2$) and 8 \AA{} ($k = 1$ eV/\AA${}^2$) to prevent pushing the formed S-O/F bond into the layer and to avoid exploration of non-interacting regions and artifacts due to periodic boundary conditions. Gaussians of width 0.1 \AA{} and height of 0.03 eV ($\sim k_B T$ to avoid trapping the CV) are deposited every 100 fs (which allows a few thermal motions of O/F in its potential well). A bias-factor of $\gamma = 100$ was chosen $\approx E_{desorb} / k_B T$ assuming a desorption barrier $E_{desorb} \sim 2-4$ eV and $T = 300$ K. Convergence analysis was performed according to the official guidelines (see Figures \ref{fig:F1S1}c and \ref{fig:F1S1}f). Convergence times, shown in the titles of Figures \ref{fig:F1S1}d and \ref{fig:F1S1}e, are $\sim 20-30$ ps. Each metaD calculation was run for at least 100 ps.

\section{Figures}

\begin{figure}
  \centering
  \includegraphics[width=1\linewidth]{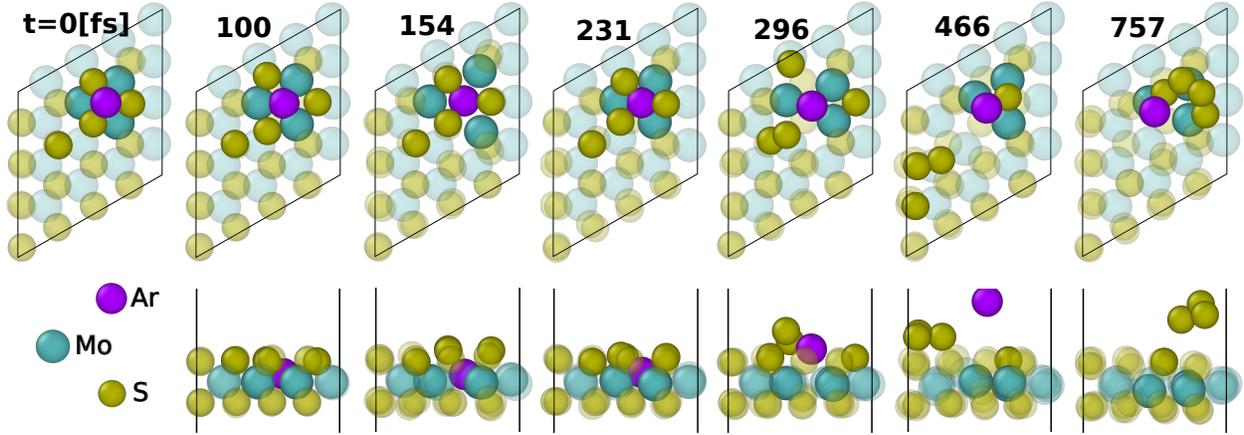}
  \caption{ Typical sputtering of pristine \ce{MoS2} at $\EArPerp = 34$ eV. Video is given in video V2. (Top) plane view, (Bottom) side view. Gray numbers indicate timestamps in [fs]. The \ce{Ar} atom has enough energy to penetrate through the top S layer and push apart Mo atoms (154 fs). These Mo atoms later push the \ce{Ar} atom back out of the lattice as follows: First (231 fs), Ar remains located “within” the Mo layer, but the Mo atoms return from their displaced state. Then, the rebounding Molybdenum atoms strongly drive the \ce{Ar} atom forward, which then collides with the top-layer S atoms at 296 fs. The displaced S-s subsequently combine to form an \ce{S3} species that then depart. Using a search similar to one described in Figure \ref{fig:F2S5}, we showed that this mechanism is the most energy efficient.}
  \label{fig:F3S1}
\end{figure}

\begin{figure}
  \centering
  \includegraphics[width=1\linewidth]{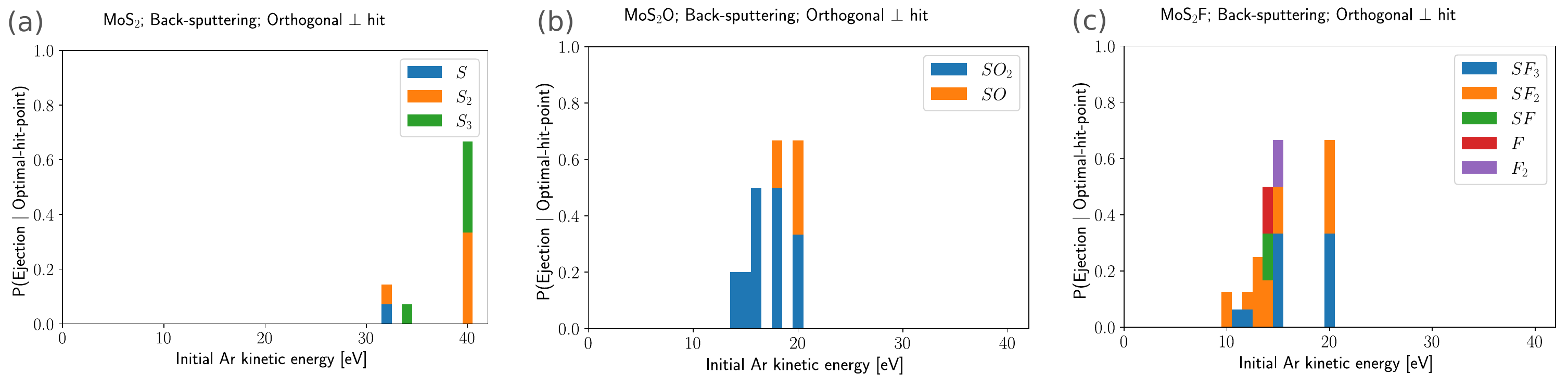}
  \caption{Breakdown of sputtering probabilities from Figure \ref{fig:F2} by products. This further illustrates a cleaner sputtering nature for \ce{MoS2O} compared to more noisy sputtering observed for \ce{MoS2F}.}
  \label{fig:F2S1}
\end{figure}

\begin{figure}
  \centering
  \includegraphics[width=1\linewidth]{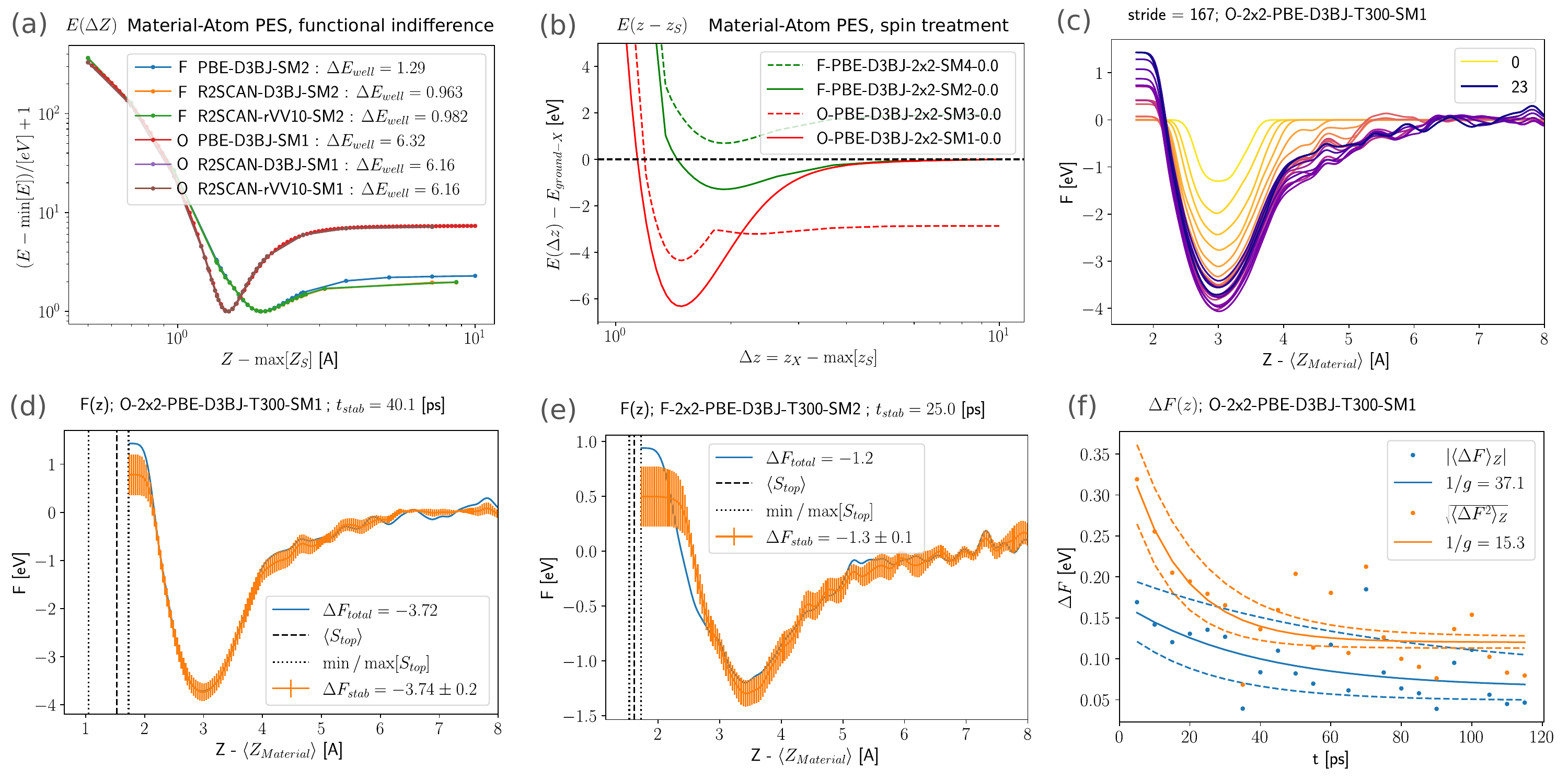}
  \caption{Calculations for O and F adsorption on \ce{MoS2}. (a) Potential energy profiles for placing an Oxygen (red) or Fluorine (green) atom at different distances from the top sulfur layer of a geometry-optimized \ce{MoS2} 2x2 single layer. The close agreement between R2SCAN and PBE curves indicates that using a mGGA is not essential. Similarly, the R2SCAN-rVV10 and R2SCAN-D3BJ curves suggest no significant advantage of rVV10 over D3BJ dispersion correction. (b) Spin-states analysis in a setup similar to A. Oxygen is red, Fluorine is green. Abrupt changes at small $r$ are because we focused on plotting near-the-well points. Solid lines are what is used in AIMD collision simulations. Spin-excited (quintuplet) F-case is higher than doublet everywhere. The O-case triplet becomes the ground state at $r > 2.15$ \AA{}. However, O is clearly a singlet when adsorbed, and there are no metastable states near $r=2.15$ \AA{}, which suggests O will be near $r=2.15$ \AA{} only for short times, making the spin-flips unlikely. The triplet PES is useful to confirm the previously reported dissociation barrier of $\sim 1.3$ eV. (c) Free-energy profile of an oxygen atom as a function of its distances from the center-of-mass of a 2x2 \ce{MoS2} single layer. Dynamics are run with unconstrained spin to test whether dynamical spin polarization affects the profile near the singlet-triplet degeneracy region. (d) Same as (c), but for a fluorine atom. (e) A typical time-evolution of meta-dynamics free-energy profile. (f) A typical convergence analysis for meta-dynamics.}
  \label{fig:F1S1}
\end{figure}

\begin{figure}
  \centering
  \includegraphics[width=1\linewidth]{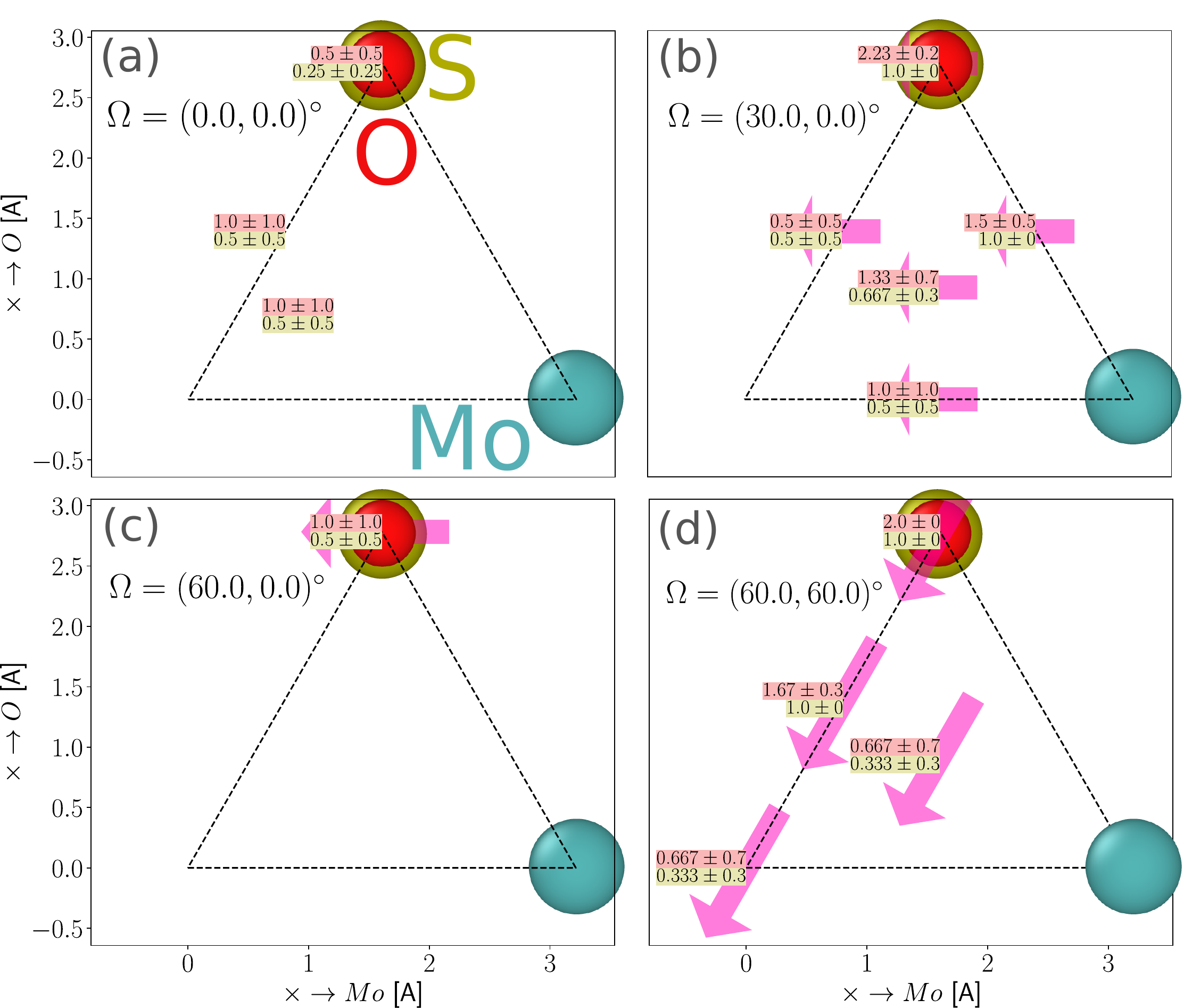}
  \caption{Finding the most damage-susceptible point of \ce{MoS2O}. Impacts are performed at $\EAr = 15$ eV. The triangle on each scheme corresponds to the purple triangle in Figure \ref{fig:F1}d, oriented according to the \ce{S} and \ce{Mo} atom labels. Thick purple arrows indicate projections of the initial \ce{Ar} velocity. Different hit-points and impact angles are surveyed (a-d). The only hit-point that always results in sulfur sputtering is directly above sulfur, corresponding to a head-on impact on oxygen in its equilibrium position.}
  \label{fig:F2S5}
\end{figure}

\begin{figure}
  \centering
  \includegraphics[width=1\linewidth]{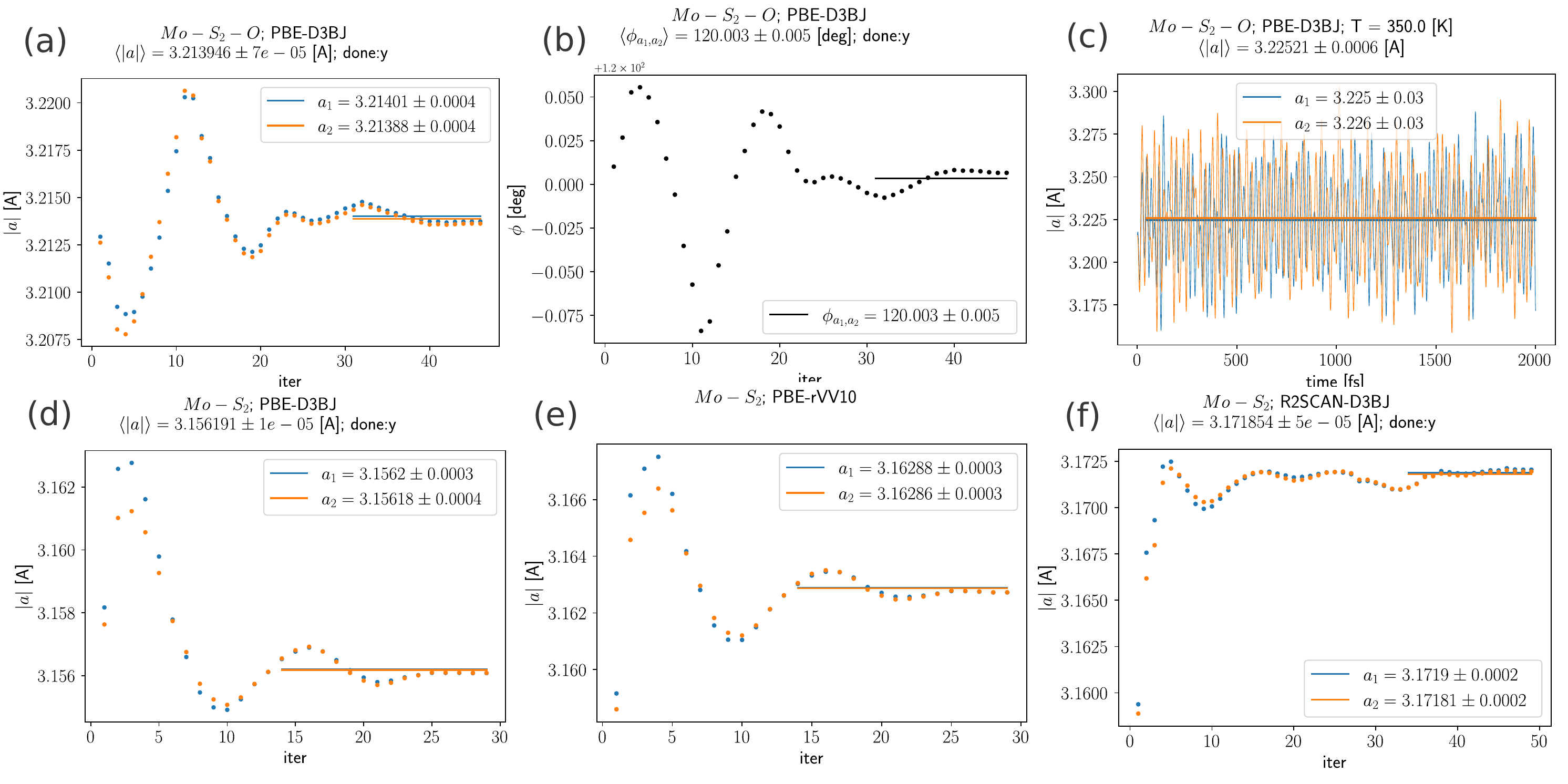}
  \caption{Equilibration before impact illustrated for \ce{MoS2O}. Geometry optimization was performed in the $NP_{xy}L_zT$ ensemble to allow the unit cell to break three-fold symmetry while preventing the collapse along the $Z$ direction. (a) Evolution of both lattice constants during optimization. (b) Evolution of the angle between the lattice constant vectors. (c) $NP_{xy}L_zT$ AIMD trajectory confirming that the ground-state optimization result from (a) is close to the lowest free-energy state. (d) Geometry optimization of \ce{MoS2}. (e) Same as (d), but including rVV10 dispersion with parameters for TMDs \cite{Peng2017, Ning2022}. The results in fact deviate even more from the experimental value of $\sim 3.15$ \AA{} \cite{Young1968}. (f) Same as (d), but using R2SCAN functional with D3 dispersion parameters optimized for TMDs \cite{Ning2022}. }
  \label{fig:F2S2}
\end{figure}

\begin{figure}
  \centering
  \includegraphics[width=1\linewidth]{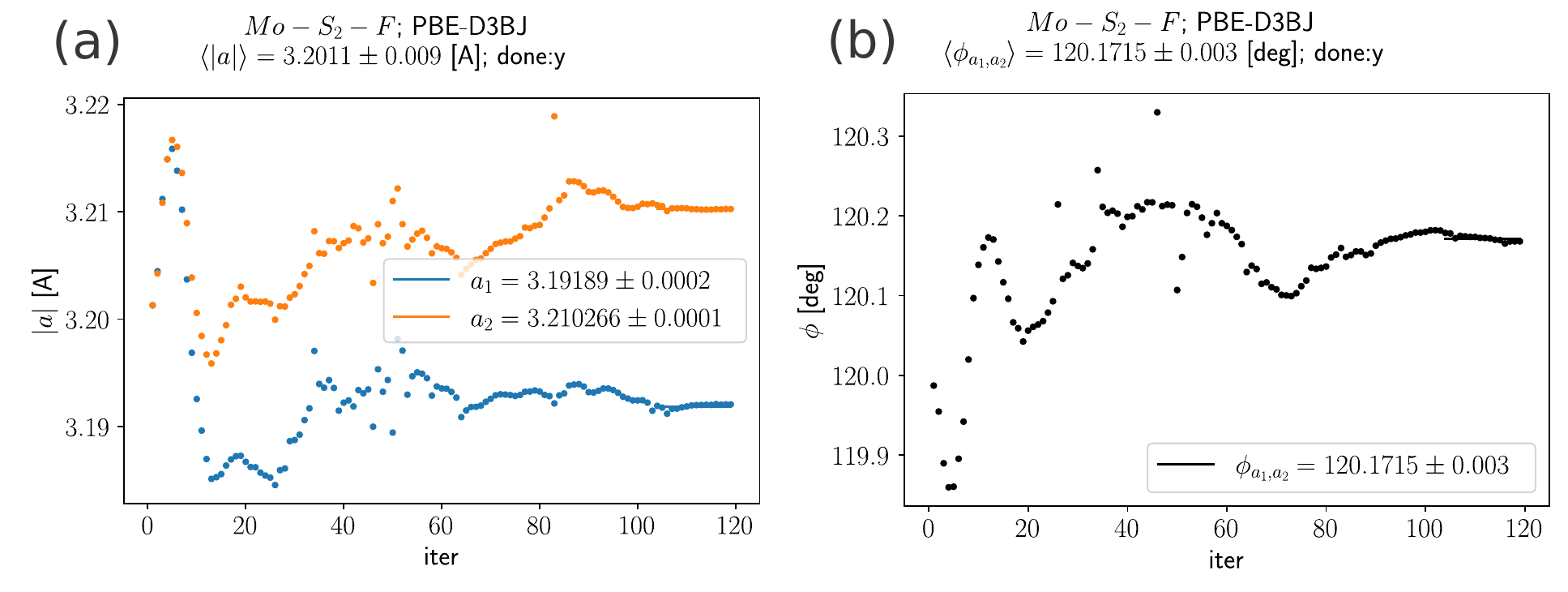}
  \caption{ Geometry optimization of \ce{MoS2F}. PBE+D3 is chosen based on its comparison to alternatives in Figure \ref{fig:F2S2}. (a) Lattice constants evolution. Spontaneous symmetry breaking is observed, as the lattice constants differ well beyond the error bars. (b) Angle between lattice constant vectors, which also deviates significantly from $120^{\circ}$ beyond the error bars.}
  \label{fig:F2S3}
\end{figure}

\begin{figure}
  \centering
  \includegraphics[width=1\linewidth]{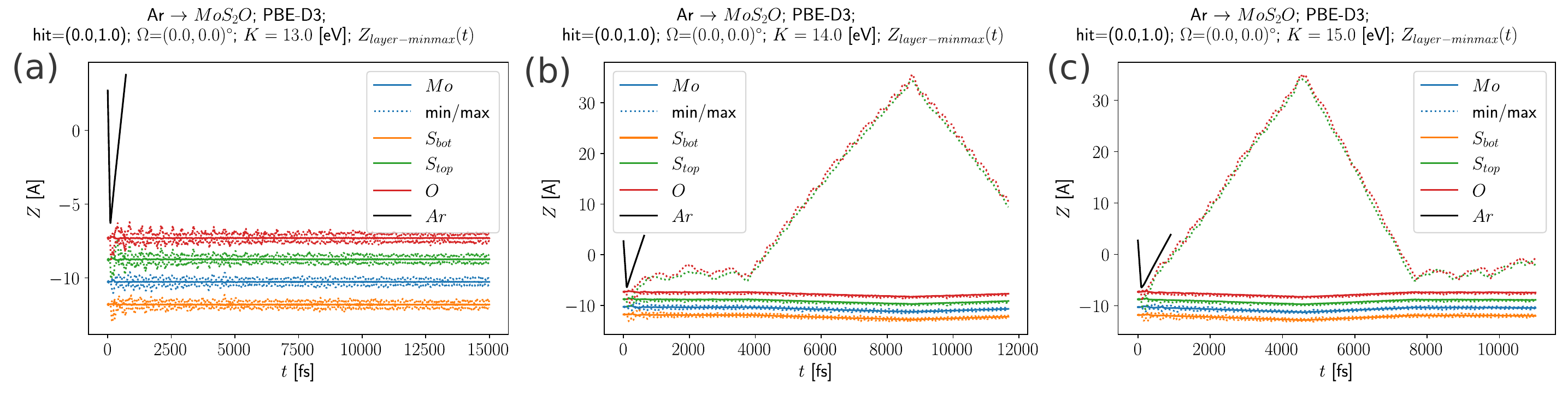}
  \caption{ Sample longer AIMD runs verifying that a 2ps simulation time used for most production results is sufficient to identify immediate sputtering events. Red, Green, Blue, and Yellow correspond to the O, $S_{top}$, Mo, and $S_{bot}$ layers, respectively. Dotted lines indicate the maximum and minimum Z-coordinates within each layer. An \ce{SO2} forms on (b) at 14 eV, but takes several ps to desorb, while an impact energy of 15 eV is sufficient for immediate sputtering. In general, removal of sputtered products involves an activation process that can occur on a much longer timescale. However, such products can also form through mechanisms other than \ce{Ar} impacts and would require a different analysis, which is beyond the scope of this work.}
  \label{fig:F2S4}
\end{figure}

\begin{figure}
  \centering
  \includegraphics[width=1\linewidth]{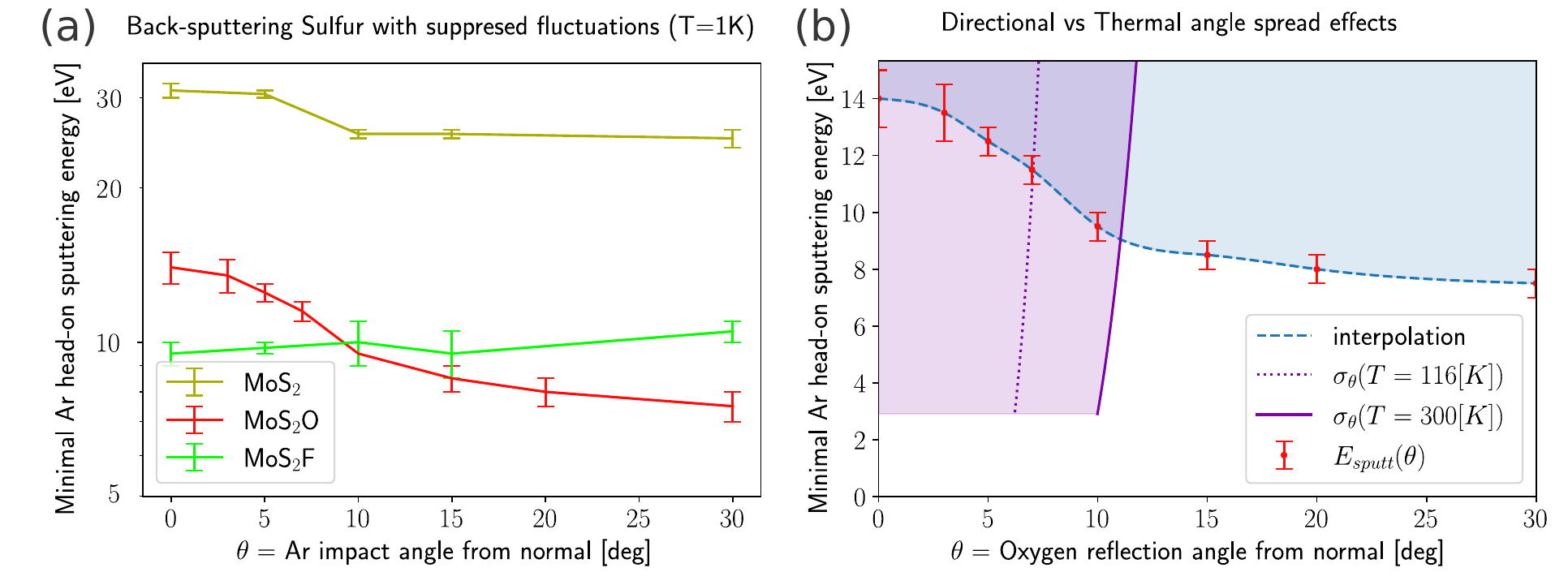}
  \caption{(a) Angular dependence of the sputtering threshold energy $\Esputt$ for pristine (dark yellow), fluorinated (light green) and oxygenated (red) \ce{MoS2} under normal hits. Such fluctuations are suppressed by setting $T = 1$ K. Each curve is obtained after optimizing the in-plane angle $\varphi$ to minimize $\Esputt$, as shown for \ce{MoS2O} in Figure \ref{fig:F4S2}. Simulations at $\theta = 45^{\circ}$ for \ce{MoS2} and \ce{MoS2O} yielded $E_{MoS_2}(45^{\circ}) > 35$ eV and $E_{MoS_2O}(45^{\circ}) > 14$ eV. (b) Sputtering possibility diagram for \ce{MoS2O}. Red datapoints indicate the \ce{Ar} energy threshold for sulfur ejections in non-normal head-on \ce{Ar}-\ce{O} collisions with suppressed thermal fluctuations (T=1 K). The blue dashed line shows a smooth (cubic) interpolation of the red data, and the blue shaded region marks where sputtering occurs. The solid purple line and corresponding shade represent the range of reflection angles $\theta$ that an O atom is likely to acquire after a \ce{Ar} normal impact at $T = 300$ K, while the purple-dotted line shows the thermal spread $\theta_T$ accessible at $T = 116$ K. Overlap between the blue and purple regions indicates where sputtering from normal impacts is expected.}
  \label{fig:F4S0}
\end{figure}

\begin{figure}
  \centering
  \includegraphics[width=0.65\linewidth]{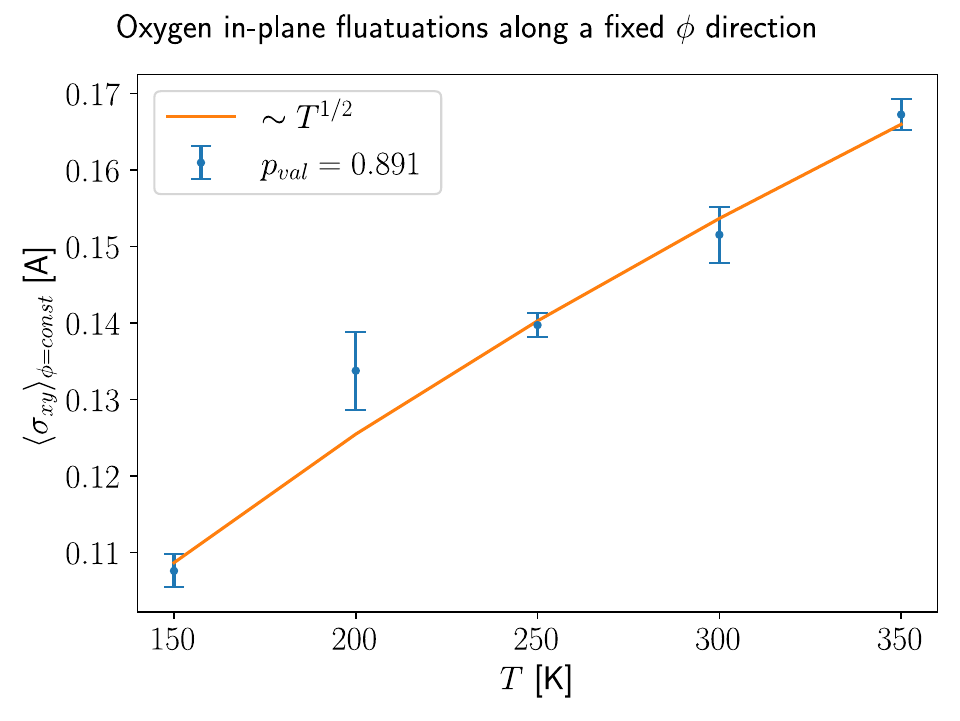}
  \caption{ Magnitude of in-plane thermal fluctuations of oxygen atoms adsorbed on \ce{MoS2}. The $\sim T^{1/2}$ scaling follows from approximating the adsorbed atoms as confined in an approximately harmonic square well potential, which implies $\sigma_{xy}^2 \sim \langle U \rangle = \langle K \rangle \sim k_B T$ by the virial theorem. The high p-value indicates that the $T^{1/2}$ scaling provides a good description of the data. We redefine $\sigma^2_{xy, \varphi=\mathrm{const}} = (\sigma^2_x + \sigma^2_y)/2$, where the factor $1/2$ reflects that we are only interested in fluctuations along a relatively narrow range of in-plane angles $\varphi \approx \phiOpt \pm \delta \varphi$ with $\delta \varphi\sim 8^{\circ}$ (Figure \ref{fig:F4S2}b). Only this range is susceptible to damage near the sputtering threshold $\Esputt$. We also assumed that in-plane fluctuations have the same magnitude in any in-plane direction, which is probably not exactly accurate because of in-plane inhomogeneity. However, $\sigma_x$ was found to be within error-bars of $\sigma_y$ in repeated simulations, which justifies computing a simple average for $\sigma^2_{xy, \varphi=\mathrm{const} }$.}
  \label{fig:F4S1}
\end{figure}

\begin{figure}
  \centering
  \includegraphics[width=1\linewidth]{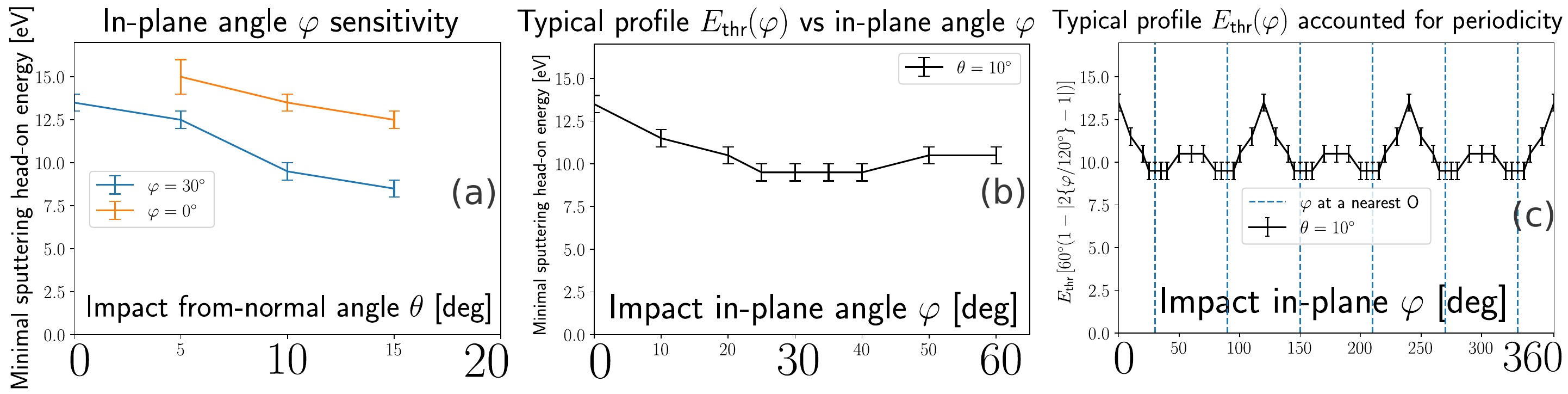}
  \caption{ Importance of the in-plane impact angle $\varphi$ for \ce{MoS2O}. (a) Clearly higher energies are required for sputtering at $\varphi = 0^{\circ}$ than at $30^{\circ}$. We note that the curves do not have to be monotonic, at least for all $\varphi \neq \phiOpt$, since pushing an O more into the wrong direction $\varphi$ can further complicate the sputtering process. We also note that the curves do not have to meet at $\theta = 0$ if fluctuations are suppressed because even an infinitesimal tilt in the wrong direction may complicate sputtering. In the presence of thermal fluctuations, the curves should meet for angles smaller than the thermal angular spread $\theta < \theta_T(\EArPerp)$. (b) Finer resolution for $\varphi$ dependence at a fixed $\theta = 10^{\circ}$. The $\theta$ is chosen big enough to make the $\theta$-dependence clearly noticeable, but not too big so the obtained dependence may be interesting for thermally induced $\theta$ spread. The minima for $\varphi \in [25; 40]^{\circ}$ corresponds to pushing the impacted O atom near-directly at a neighboring O atom, which is located at $\varphi = 30^{\circ}$. (c) The data from (b), unwrapped using the symmetry (see Figure \ref{fig:F1}). The applied symmetry is only relevant for the elementary triangle corners (the purple triangle on Figure \ref{fig:F1}). The system is less symmetric for hit-points on triangle edges and even less so inside the triangle. Blue dashed lines note the direction at all 6 neighboring O-s.  }
  \label{fig:F4S2}
\end{figure}

\begin{figure}
  \centering
  \includegraphics[width=1\linewidth]{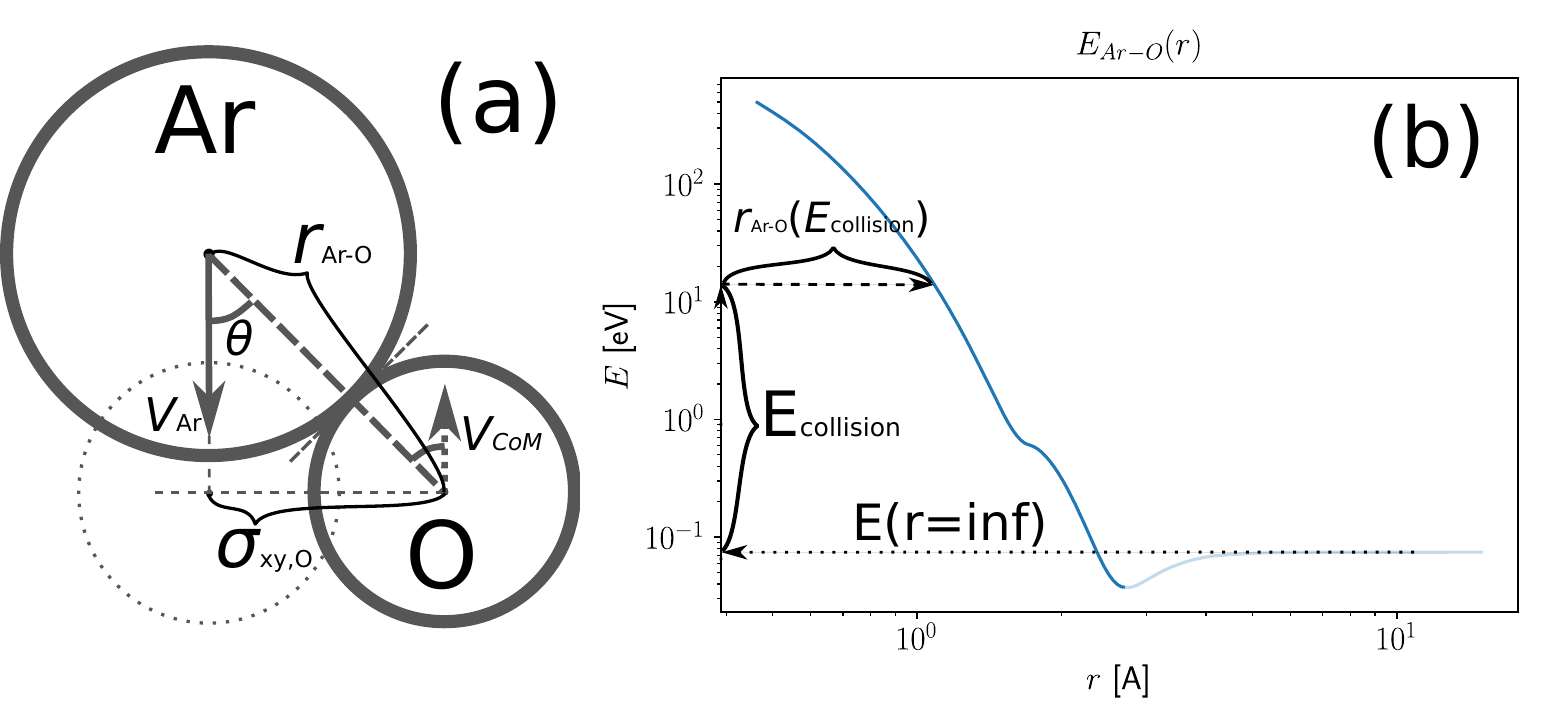}
  \caption{ Ar-O collision schematic. (a) Hard-sphere Ar-O collision model. The dotted circle is the equilibrium O position, while the O circle shifted by $\sigma_{xy,O}$ to the right represents its position at the moment of impact with \ce{Ar}. The collision plane (thin dashed line) is orthogonal to the line through the centers of \ce{Ar} and O (thick dashed). The incoming \ce{Ar} has velocity $\vAr$, however treating the collision in the CoM frame assigns the O atoms a velocity $\vCoM$ and subtracts the same $\vCoM$ from $\vAr$. Because $\sigma_{xy,O} \sim \rArO$ we get non-negligible $\theta = \texttt{arcsin}(\sigma_{xy,O} / \rArO)$. A non-head-on collision can always be decomposed into a head-on collision and a fly-by (orthogonal to the head-on collision). (b) Two-atom potential energy profile $\UArO(r)$ for Ar-O used in eq.\eqref{eq:U_rArX}. The value of $\rArO$ is found as $\UArO^{-1}(\Ecollision)$. See the Supplementary information for the derivation of $\Ecollision$.}
  \label{fig:F4S3}
\end{figure}

\clearpage



\end{suppinfo}


\bibliography{achemso-demo}

\end{document}